\documentclass[preprint2]{aastex}
\usepackage{natbib}

\slugcomment{}

\usepackage{graphicx}
\usepackage{amsmath}
\usepackage{natbib}
\usepackage{epsfig}
\usepackage{txfonts}

\begin{document}

\title{Impact of supernova dynamics on the $\nu p$-process}

\author{A.~Arcones\altaffilmark{1} }
\affil{Department of Physics, University of Basel, CH-4056 Basel, Switzerland}
\email{a.arcones@unibas.ch}
\author{C.~Fr\"ohlich}
\affil{Department of Physics, North Carolina State University, Raleigh NC 27695}
\email{cfrohli@ncsu.edu}
\and
\author{G.~Mart\'inez-Pinedo}
\affil{Institut f\"ur Kernphysik, Technische Universit\"at Darmstadt, D-64289 Darmstadt, Germany\\
GSI Helmholtzzentrum f\"ur Schwerionenforschung, D-64291 Darmstadt, Germany }

\altaffiltext{1}{Feodor Lynen Fellow, Alexander von Humboldt Foundation}

\begin{abstract}
  We study the impact of the late time dynamical evolution of ejecta
  from core-collapse supernovae on $\nu p$-process nucleosynthesis.
  Our results are based on hydrodynamical simulations of neutrino wind
  ejecta.  Motivated by recent two-dimensional wind simulations, we
  vary the dynamical evolution during the $\nu p$-process and show
  that final abundances strongly depend on the temperature
  evolution. When the expansion is very fast, there is not enough time
  for antineutrino absorption on protons to produce enough neutrons to
  overcome the $\beta^+$-decay waiting points and no heavy elements
  beyond $A=64$ are produced.  The wind termination shock or reverse
  shock dramatically reduces the expansion speed of the ejecta.  This
  extends the period during which matter remains at relatively high
  temperatures and is exposed to high neutrino fluxes, thus allowing
  for further $(p,\gamma)$ and $(n,p)$ reactions to occur and to
  synthesize elements beyond iron.  We find that the $\nu p$-process
  starts to efficiently produce heavy elements only when the
  temperature drops below $\sim 3$~GK.  At higher temperatures, due to the
  low alpha separation energy of $^{60}$Zn ($S_{\alpha} = 2.7$~MeV)
  the reaction $^{59}$Cu$(p,\alpha)$$^{56}$Ni is faster than the
  reaction $^{59}$Cu$(p,\gamma)$$^{60}$Zn.  This results in the closed
  NiCu cycle that we identify and discuss here for the first time.  We
  also investigate the late phase of the $\nu p$-process when the
  temperatures become too low to maintain proton captures.  Depending
  on the late neutron density, the evolution to stability is dominated
  by $\beta^+$ decays or by $(n,\gamma)$ reactions.  In the latter
  case, the matter flow can even reach the neutron-rich side of
  stability and the isotopic composition of a given element is then
  dominated by neutron-rich isotopes.
\end{abstract}


\section{Introduction}
\label{sec:intro}

Neutrino-driven winds from core-collapse supernova explosions
contribute to the synthesis of elements beyond iron. After the
explosion, the hot proto-neutron star cools emitting neutrinos.  These
neutrinos interact with the stellar matter and deposit energy in the
outer layers of the proto-neutron star leading to a supersonic outflow
known as neutrino-driven wind \cite[]{duncan.shapiro.wasserman:1986}.
Although neutrino-driven winds were considered the site where heavy
elements are produced by the r-process \cite[]{Woosley94}, recent
simulations \cite[]{arcones.janka.scheck:2007, Huedepohl.etal:2010,
  Fischer.etal:2010, Roberts.etal:2010} cannot reproduce the extreme
conditions required for producing heavy r-process elements \cite[see
e.g.,][]{hoffman.woosley.qian:1997, Otsuki.Tagoshi.ea:2000,
  Thompson.Burrows.Meyer:2001}. The wind entropy is too low (less than
$100\,k_{\mathrm{B}}/\mathrm{nuc}$) and, even more significant, the
ejecta is proton rich (the electron fraction $Y_e$ remains above 0.5
during seconds, see \cite{Huedepohl.etal:2010,
  Fischer.etal:2010}). Even if the r-process does not take place in
every neutrino-driven wind, lighter heavy elements (e.g., Sr, Y, Zr)
can be synthesized in this environment as suggested by
\cite{Qian.Wasserburg:2001}. In proton-rich conditions
\cite{Froehlich06} showed that elements beyond $^{64}$Ge can be
synthesized. \cite{Wanajo.Janka.Mueller:2011} found Sr, Y, Zr in small
pockets of neutron rich material ejected after the explosion of low
mass progenitors. Recently, \cite{Arcones.Montes:2011} performed a
systematic nucleosynthesis study that strongly supports the production
of lighter heavy elements in proton- and neutron-rich neutrino-driven
winds.

In proton-rich winds, charged particle reactions (alpha and proton
captures) build nuclei up to $^{56}$Ni and even up to $^{64}$Ge once the
temperature drops below 3~GK. Due to their long beta-decay lifetimes
and low thresholds for proton capture, the nuclei $^{56}$Ni and
$^{64}$Ge act as bottlenecks that inhibit the production of heavier
elements. In the $\nu p$-process, their decay is sped up by $(n,p)$
reactions, with the neutrons produced by antineutrino absorption on
the abundant free protons. This allows for the production of elements
beyond iron and may explain the origin of light p-nuclei
\cite[]{Froehlich06, Pruet.Hoffman.ea:2006, Wanajo:2006,
  Wanajo.etal:2011}.  The synthesis of elements by the $\nu p$-process
depends thus on neutrino spectra and luminosities but also on the
dynamical evolution as matter expands through the slow, early
supernova ejecta.  This produces a wind termination shock or reverse
shock where kinetic energy is transformed into internal energy
\cite[]{arcones.janka.scheck:2007}. The reverse shock has a big impact
on the nucleosynthesis because temperature and density increase and
the expansion is strongly decelerated. This hydrodynamical feature has
been extensively studied for r-process nucleosynthesis
\cite[]{Qian.Woosley:1996, Sumiyoshi00, Wanajo:2007, Panov.Janka:2009,
  Arcones.MartinezPinedo:2011}. Recently, \cite{Wanajo.etal:2011} have
also explored the relevance of the reverse shock on the $\nu
p$-process. Motivated by their work and by new 2D hydrodynamical
simulations of the neutrino-driven wind \cite[]{Arcones.Janka:2011},
we investigate here the wind termination shock to gain further insights on the
dynamical evolution relevant for the $\nu p$-process.

Here we use a trajectory from hydrodynamical simulations
(Sect.~\ref{sec:dynamics}) combined with a complete nucleosynthesis
network (Sect.~\ref{sec:netw}).  In Sect.~\ref{sec:results} we present
our results where we analyze the impact of the wind termination
temperature (Sect.~\ref{sec:constT}), the temperature jump at the
reverse shock (Sect. \ref{sec:rs}), and the influence of late
temperature evolution and the decay to stability
(Sect~\ref{sec:long}).  Our conclusions are summarized in
Sect.~\ref{sec:conclusions}.

\section{Long-time dynamical evolution}
\label{sec:dynamics}
Our study is based on one trajectory obtained from hydrodynamical
simulations by \cite{arcones.janka.scheck:2007}. These simulations
follow the supernova explosion and the subsequent neutrino-driven
wind. 

\begin{figure}[!htb]
  \includegraphics[width=\linewidth]{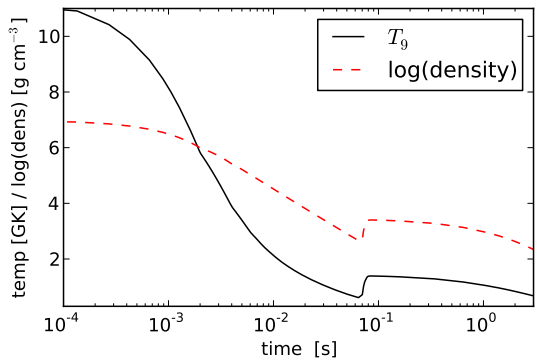}\\
  \caption{Evolution of density and temperature along a trajectory
    ejected at 2s after bounce of model M15l1r1 in
    \cite{arcones.janka.scheck:2007}. The reverse shock is at
    $T_{\mathrm{wt}} \approx 0.6$~GK and produces a jump in density
    and temperature and a sudden decrease of the expansion velocity.}
  \label{fig:trajectory1D}
\end{figure}

Figure~\ref{fig:trajectory1D} presents the evolution of density and
temperature for the chosen trajectory. In the following, we always use
the same initial evolution and modify it at temperatures 
below 3~GK.  We assume the wind terminates at a temperature
$T_{\mathrm{wt}}$ and the evolution thereafter is varied
using the prescription introduced by
\cite{Arcones.MartinezPinedo:2011} and motivated by 1D and 2D
hydrodynamical simulations
\cite[]{arcones.janka.scheck:2007,Arcones.Janka:2011}. The
thermodynamical conditions at the termination of the wind depend on
both the intrinsic properties of the wind, like its velocity and mass
outflow rate, and on the pressure of the slow moving ejecta. The wind
properties are determined by the neutrino energies and luminosities
and by the neutron star mass and radius~\cite[see
e.g.,][]{Qian.Woosley:1996}. These conditions depend mainly on the
nuclear equation of state and are similar for different progenitors
\cite[]{arcones.janka.scheck:2007,Fischer.etal:2010}. The properties
of the slow supernova ejecta are related to the progenitor and become very
anisotropic during the evolution
\cite[]{Arcones.Janka:2011}.

In our calculations, we use a simple model that reproduces the main
features seen in hydrodynamical simulations. Once the wind reaches the
early supernova ejecta we use the Rankine-Hugoniot conditions to
determine the behavior of temperature, density, and velocity.  When
the wind moves supersonically these quantities jump at the reverse shock. 
The evolution after such discontinuity is
determined assuming constant density and temperature during a time
$\Delta t$. As the mass outflow is constant ($\dot{M} = 4\pi r^2 v
\rho$) the velocity drops as $v \propto r^{-2}$. At later times, the
velocity stays constant and the density decreases as $\rho \propto
r^{-2}$ (see \cite{Arcones.MartinezPinedo:2011} for more details). To
account for the differences found in the late evolution in
two-dimensional simulations, we will use different values of $\Delta
t$ and explore the impact on the nucleosynthesis.

\section{Nucleosynthesis network}
\label{sec:netw}

The evolution of the composition is calculated using a full reaction
network \cite[]{Froehlich06}.  All calculations start when the
temperature drops below 10~GK and are followed until the temperature
reaches 0.01~GK.  We assume the initial composition to be determined
from nuclear statistical equilibrium for a fixed electron fraction
$Y_e = 0.52$.  The reaction network includes 1869~nuclei from free
nucleons to dysprosium ($Z=66$).  Neutral and charged particle
reactions are taken from the REACLIB compilation, containing
theoretical statistical-model rates by \cite{Rauscher.Thielemann:2000}
(NON-SMOKER) and experimental rates by \cite{Angulo.Arnould.ea:1999}.
The theoretical weak interactions rates are taken from
\cite{Langanke.Martinez-Pinedo:2001} and
\cite{Fuller.Fowler.Newman:1982a}.  Where available, experimental
beta-decay rates are used from NuDat2\footnote{"National Nuclear Data
  Center, information extracted from the NuDat 2 database",
  http://www.nndc.bnl.gov/nudat2/} supplemented by theoretical
beta-decay rates~\cite[]{Moeller.Pfeiffer.Kratz:2003}.  Neutrino
absorption on nucleons and nuclei is also taken into account.  We
assume a constant neutrino luminosity for neutrinos and anti-neutrinos
($ L_{\nu_e}= L_{\bar{\nu}_e} = 5 \times 10^{51}$~erg), and
Fermi-Dirac spectra with a temperature consistent with the
hydrodynamical simulations.

In order to understand the nucleosynthesis evolution, it is useful to
look at the time variations of the mean lifetimes for
$\beta^+$-decays, $(n,\gamma)$, $(p,\gamma)$, $(n,p)$, $(\gamma,n)$,
and $(\gamma,p)$ reactions.  These mean lifetimes (or average
timescales) are defined as

\begin{equation}
\tau_{\beta^+} \equiv \left[ \frac{1}{Y_h} \sum_{Z>2,A} \lambda_{\beta^+}(Z,A) Y(Z,A) \right] ^{-1},
\end{equation}

\begin{equation}
\tau_{n \gamma} \equiv \left[ \frac{\rho Y_n}{Y_h} N_A \sum_{Z>2,A} \langle \sigma v \rangle_{(n,\gamma)}(Z,A) Y(Z,A) \right]^{-1},
\end{equation}

\begin{equation}
\tau_{p \gamma} \equiv \left[ \frac{\rho Y_p}{Y_h}  N_A \sum_{Z>2,A} \langle \sigma v \rangle_{(p,\gamma)}(Z,A) Y(Z,A) \right]^{-1},
\end{equation}

\begin{equation}
\tau_{np} \equiv \left[ \frac{\rho Y_n}{Y_h} N_A \sum_{Z>2,A} \langle \sigma v \rangle_{(n,p)}(Z,A) Y(Z,A) \right]^{-1},
\end{equation}

\begin{equation}
\tau_{\gamma n} \equiv \left[ \frac{1}{Y_h} \sum_{Z>2,A} \lambda_{\gamma n}(Z,A) Y(Z,A) \right]^{-1},
\end{equation}
and
\begin{equation}
\tau_{\gamma p} \equiv \left[ \frac{1}{Y_h} \sum_{Z>2,A} \lambda_{\gamma p}(Z,A) Y(Z,A) \right]^{-1},
\end{equation}
where $\lambda_{\beta^+}(Z,A)$ denotes the decay rate of nucleus
$(Z,A)$, $\langle \sigma v \rangle_{x}(Z,A)$ the reaction rate for
process $x$ on nucleus $(Z,A)$, and $\lambda_{\gamma n}(Z,A)$, and
$\lambda_{\gamma p}(Z,A)$ the photo-dissociation reaction rates with
emission of neutron and proton, respectively.  In these equations,
$N_A$ is Avogadro number. The average is
taken over the heavy nuclei ($Z>2$):
\begin{equation}
Y_h = \sum_{Z>2,A} Y(Z,A).
\end{equation}

Another useful quantity to understand the nucleosynthesis evolution is
the reaction flow between two nuclei $i$ and $j$ defined as
\begin{equation}
F_{ij} \equiv \dot{Y}(i \rightarrow j)- \dot{Y}(j \rightarrow i)\, .
\end{equation}
The quantity $\dot{Y}(i \rightarrow j)$ denotes the change in
abundance of nucleus $i$ due to all reactions connecting nucleus $i$
with nucleus $j$. The largest reaction flows at each time step
indicate the key reactions and thus the nucleosynthesis path.

\section{Results}
\label{sec:results}

\subsection{Constant temperature}
\label{sec:constT}
\begin{figure}[!htb]
  \begin{tabular}{c}
    \includegraphics[width=\linewidth]{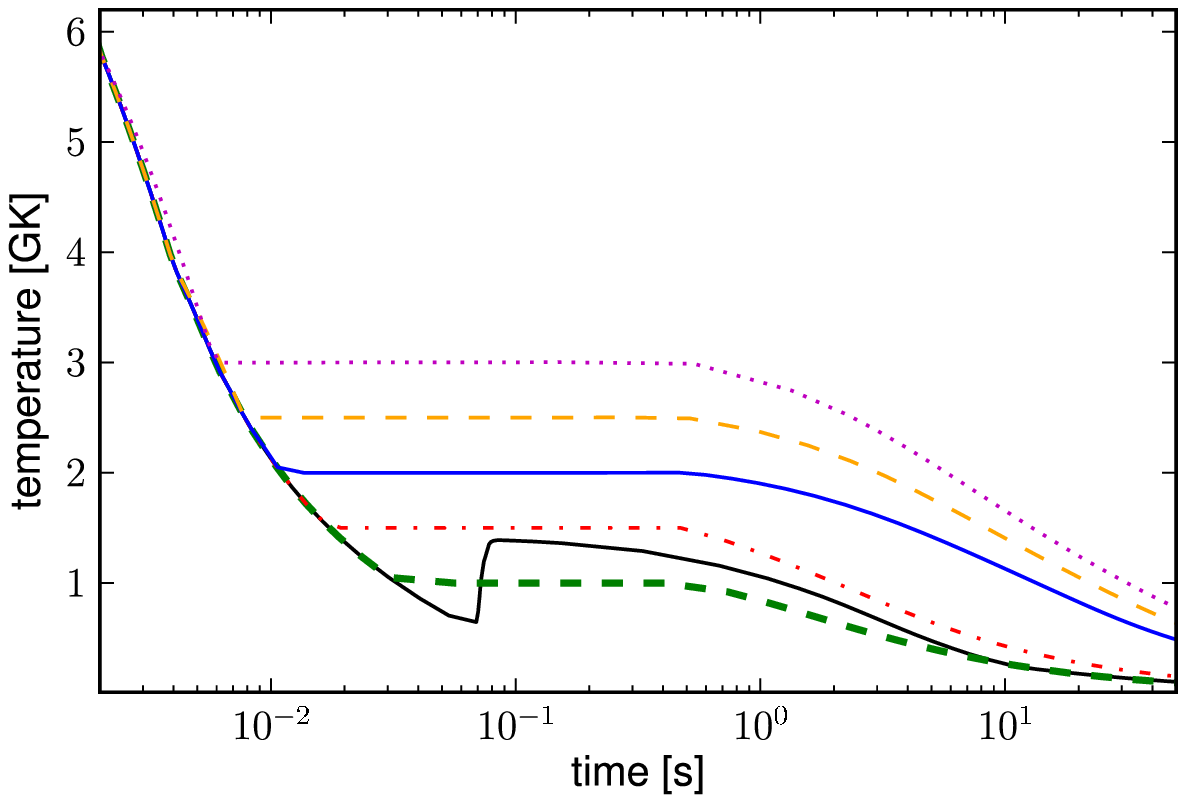}\\
    \includegraphics[width=\linewidth]{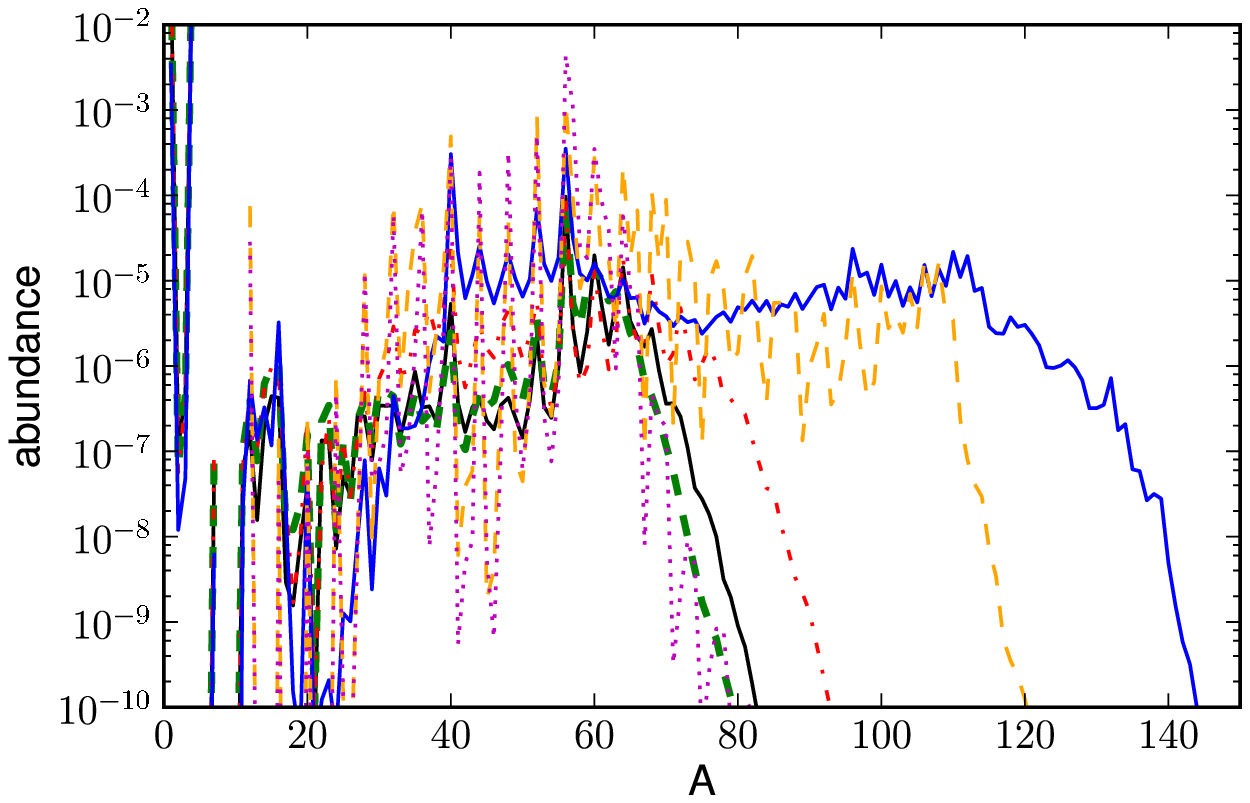}\\
    \includegraphics[width=\linewidth]{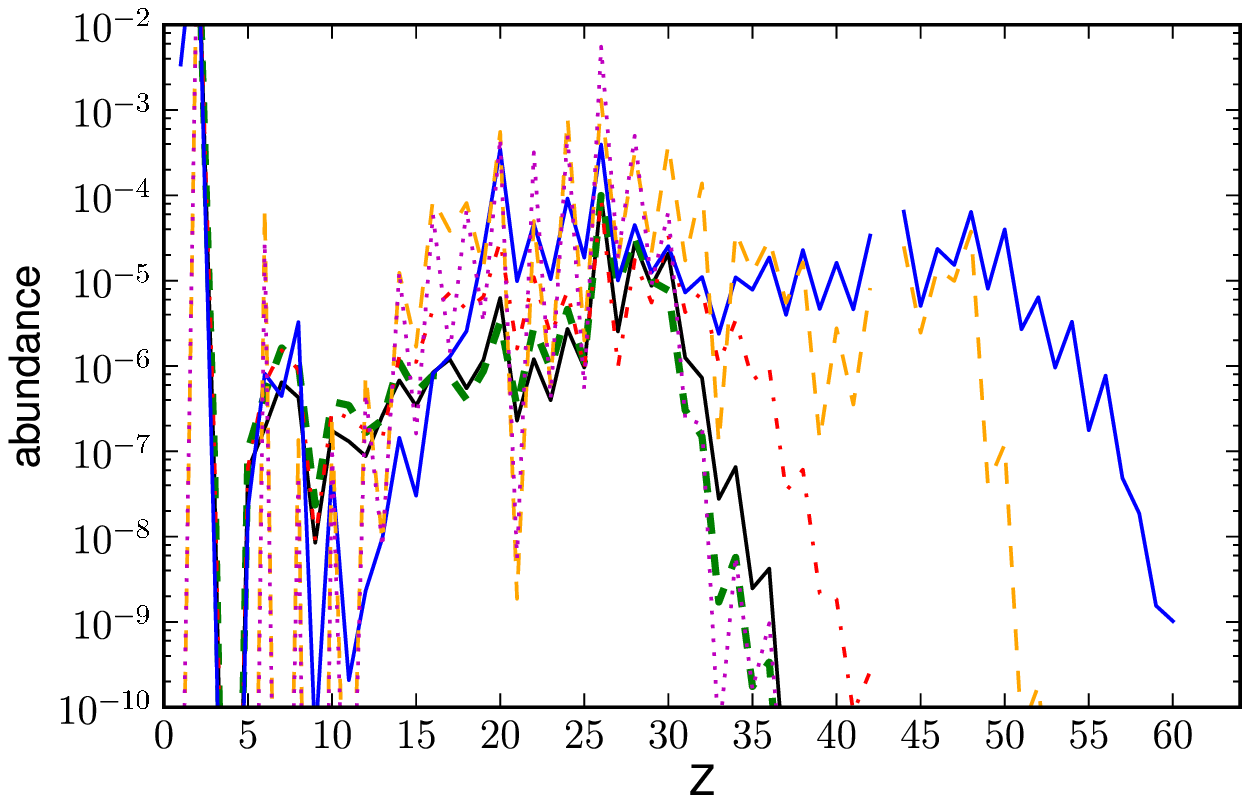}
  \end{tabular}
  \caption{Different temperature evolutions are shown in the upper
    panel. The solid black line corresponds to the original
    trajectory.  The abundances for these evolutions are presented
    (with same colors and line styles) in the middle and bottom panels
    versus mass and proton numbers, respectively.}
  \label{fig:tempconst}
\end{figure}

\begin{figure}[!htb]
 \includegraphics[width=\linewidth]{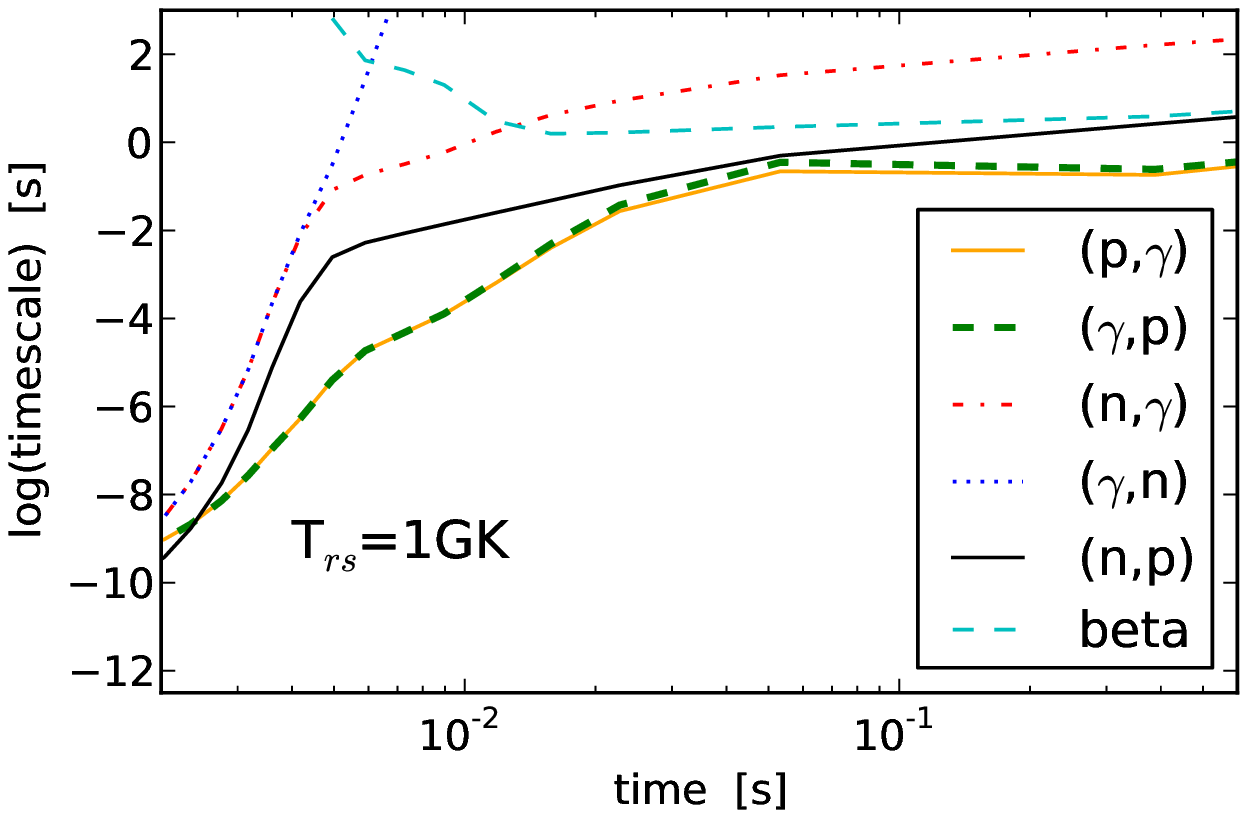}
 \includegraphics[width=\linewidth]{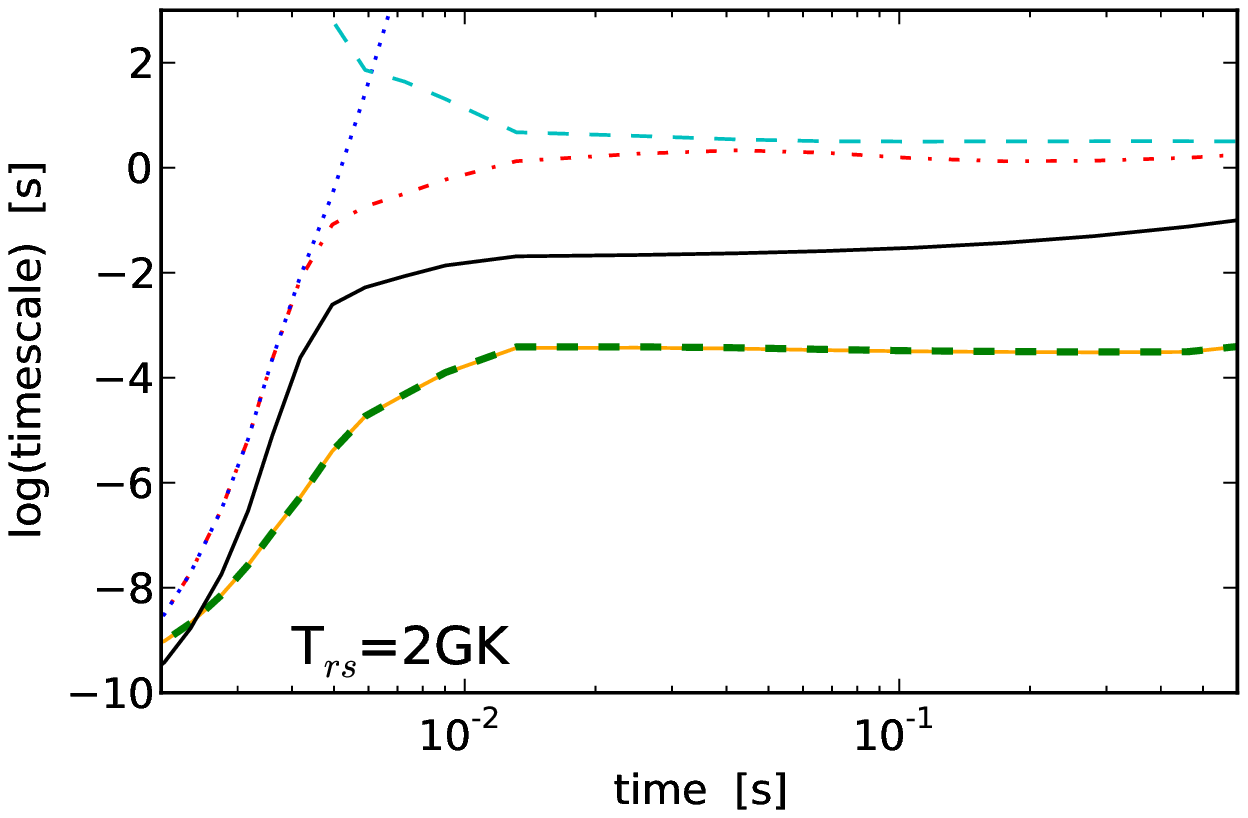}
 \includegraphics[width=\linewidth]{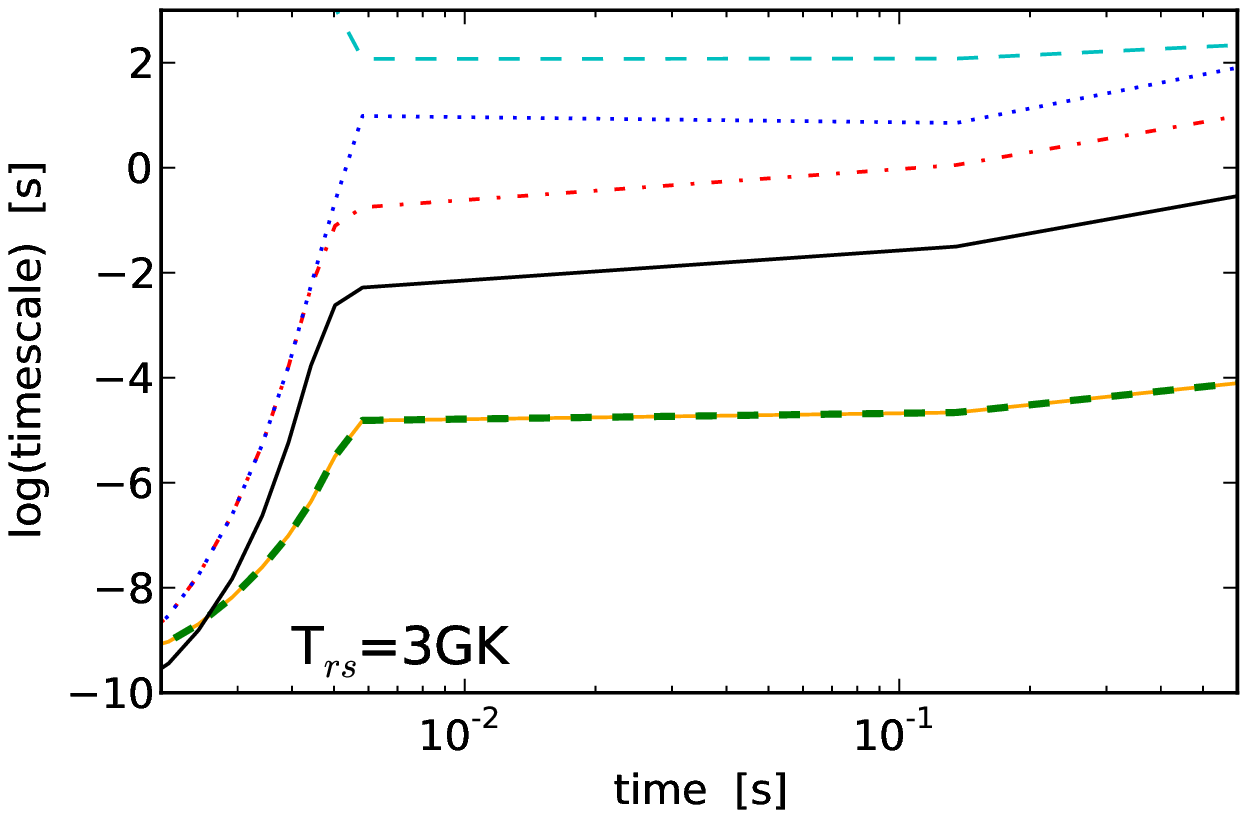}
 \caption{Averaged time scales of the most relevant reactions
   indicated in the labels. The three panels correspond to the wind
   termination at 1, 2, and 3~GK from the top to the bottom.}
  \label{fig:timescale}
\end{figure}

We study here the impact of the wind termination varying the
temperature at which it occurs.  We assume that the velocity is
subsonic and hence there is no jump in temperature, density, and
velocity.  The different temperature evolutions and their
nucleosynthesis are shown in Fig.~\ref{fig:tempconst}. The original
trajectory (solid black line) is also included for completeness and
comparison purposes. In the modified trajectories the wind termination
is at $T_{\mathrm{wt}}=$~1, 1.5, 2, 2.5, and 3~GK. The resulting
abundances change significantly with temperature $T_{\mathrm{wt}}$. 
There is an optimal $T_{\mathrm{wt}}$ around $2$~GK for producing 
heavier elements \cite[]{Wanajo.etal:2011}. Moreover, the nucleosynthesis 
evolution is very different for temperatures higher and lower than this 
optimal temperature. This can be seen in Fig.~\ref{fig:timescale} where the 
relevant averaged timescales (see Sect.~\ref{sec:netw}) are shown for
the trajectories with $T_{\mathrm{wt}}=$~1,~2,~3~GK.  Initially, there
is $(p,\gamma)-(\gamma,p)$ equilibrium in all cases.  The timescales
for the $(p,\gamma)$ and $(\gamma,p)$ reactions are much shorter when
the wind termination occurs at higher temperatures (see bottom panel of
Figure \ref{fig:timescale}).

For the wind termination at low temperatures, $T_{\mathrm{wt}}=$~1,
1.5~GK, only elements up to Germanium are produced in substantial
amounts.  In these evolutions matter expands very fast. Therefore, it
only stays for a short time in the temperature rage of $1.5-3$~GK where
charged-particle reactions can effectively synthesize heavier
nuclei. In addition, matter rapidly reaches large radii where the
antineutrino flux is rather low. Similar conditions are found in
explosions of low mass progenitors where the expansion of the
supernova ejecta is very fast \cite[]{Hoffman.Mueller.Janka:2008,
  Wanajo.Nomoto.ea:2009,Roberts.etal:2010, Wanajo.etal:2011}. Notice
that the wind termination at a slightly higher temperature, 1.5~GK
instead of 1~GK, allows to reach somewhat higher mass number.

\begin{figure}[!htb]
  \begin{center}
    \includegraphics[width=1\linewidth]{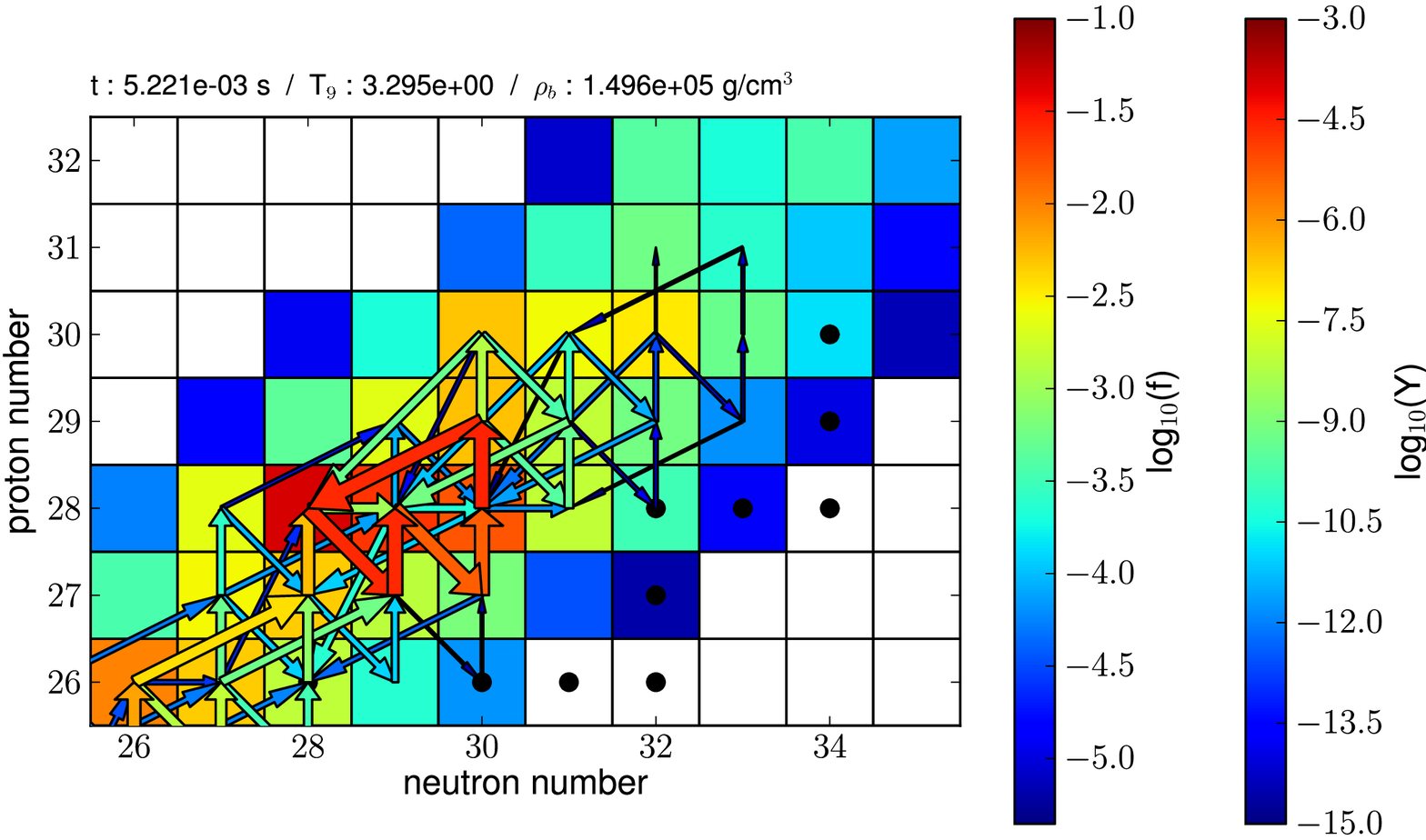}\\
    \includegraphics[width=1\linewidth]{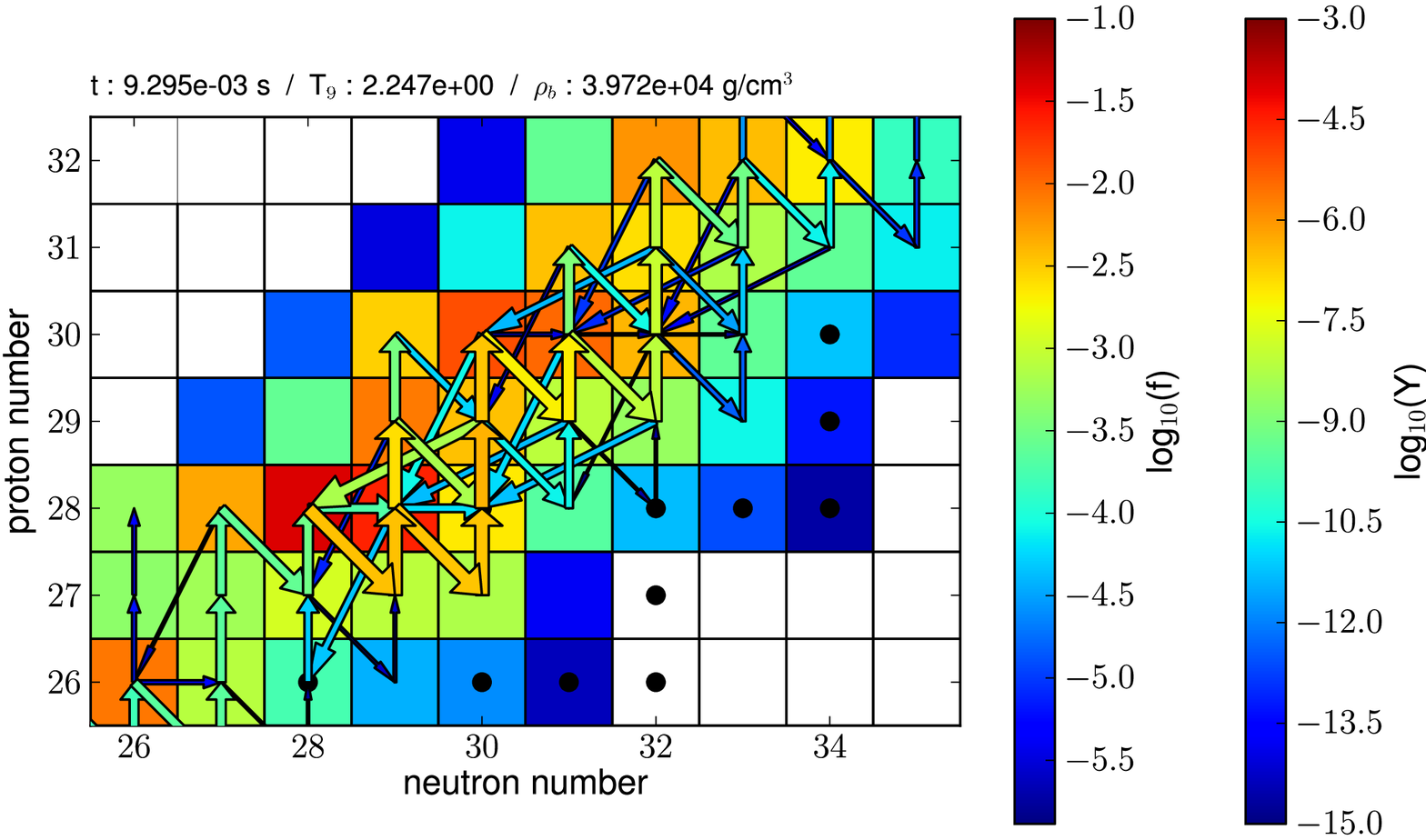}
  \end{center}
  \caption{Arrows indicate the flow of different reactions normalized
    to the strongest one. The colors and sizes of the arrows are
    proportional to the flow. Abundances are also shown with colors
    and stable nuclei are indicated with a dot. The upper panel
    corresponds to the early evolution when the temperature is around
    3.3~GK and in the bottom panel it has dropped down to
    $\sim$~2.2~GK.}
  \label{fig:flux}
\end{figure}

\begin{figure}[!htb]
 \includegraphics[width=\linewidth]{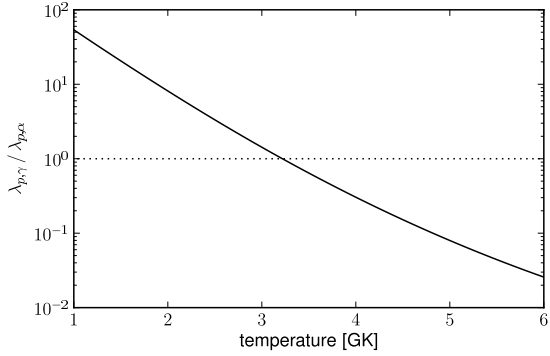}
\caption{Ratio of proton capture rates on $^{59}$Cu, with
   $\lambda_{(p,\gamma)} = \rho Y_p N_A \langle \sigma v
   \rangle_{(p,\gamma)}$ and $\lambda_{(p,\alpha)} = \rho Y_p N_A
   \langle \sigma v \rangle_{(p,\alpha)}$. At high temperatures
   $^{59}$Cu$(p,\alpha)$$^{56}$Ni dominates while for lower
   temperature $^{59}$Cu$(p,\gamma)$$^{60}$Zn dominates. Dotted line
   indicates where rates are equal.} 
  \label{fig:Cu59}
\end{figure}

For the wind termination at high temperatures, one would expect that
photo-dissociation stops the synthesis of heavier elements. In this
case, the abundance would continuously shift towards lower mass number
as temperature increases. However, in our calculations we see an
abrupt change in the abundances when the wind termination temperature
exceeds certain value. This points to a key reaction acting as a
bottleneck at high temperatures. We have identified such a reaction
using the reaction flows introduced in
Sect.~\ref{sec:netw}. Figure~\ref{fig:flux} shows these flows at two
different temperatures for the evolution with
$T_{\mathrm{wt}}=$~2~GK. The upper panel corresponds to $T \approx
3.3$~GK, and the bottom one to $T\approx 2.2$~GK, both represent
conditions before the constant temperature phase.  Notice that the
reactions that determine the nucleosynthesis flow are different at
high and low temperatures.  At high temperatures, the reaction
$^{59}$Cu$(p,\alpha)$$^{56}$Ni dominates over
$^{59}$Cu$(p,\gamma)$$^{60}$Zn while at low temperatures the contrary
is true (see Fig.~\ref{fig:Cu59}).  When the $^{59}$Cu$(p,\alpha)$
reaction dominates, the nucleosynthesis flow is confined into a closed
NiCu cycle.  A similar behavior has been found in the rp-process for
the SnSbTe cycle \cite[]{Schatz.etal:2001}.

In the calculation with $T_{\mathrm{wt}}=3$~GK, when temperatures are
still above $\sim 3.2$~GK (see Fig.~\ref{fig:Cu59}) the heaviest
nucleus produced is $^{56}$Ni due to the NiCu cycle. Once the
temperature drops, the cycle opens and the path reaches
$^{64}$Ge. However, during the high constant temperature phase, the
triple alpha reaction maintains a continuous
production of seed nuclei from light nuclei with
$A\leq7$~\cite[]{Wanajo.etal:2011}. This leads to a reduction of the
proton abundance and thus of the neutrons produced by antineutrino
absorption.  Notice that the ratio of neutrons produced per seed
nucleus is a useful guide of the strength of the $\nu
p$-process~\cite[]{Pruet.Hoffman.ea:2006}. Both effects (the NiCu
cycle and the reduced neutron-to-seed ratio) result in a
complete shutdown of $\nu p$-process nucleosynthesis.

When the wind termination temperature is very high ($\approx 3$~GK) or
very low ($\approx 1$~GK), elements beyond Germanium are hardly
synthesized. However, the final abundances are different for these two
extreme cases. When the wind termination takes place at high
temperatures ($\gtrsim3$~GK) matter accumulates in iron group nuclei due to
the NiCu cycle and to the continuous production of seed nuclei.  While
for the wind termination at low temperatures matter moves fast to
large distances where the antineutrino flux is not large enough to
produce a $\nu p$-process. In addition, the low temperatures inhibit
the production of seed nuclei. Therefore, the final proton abundance
is significantly higher in the evolution with the wind termination at
temperatures $\lesssim2$~GK.

\subsection{Reverse shock}
\label{sec:rs}

\begin{figure}[!htb]
 \includegraphics[width=\linewidth]{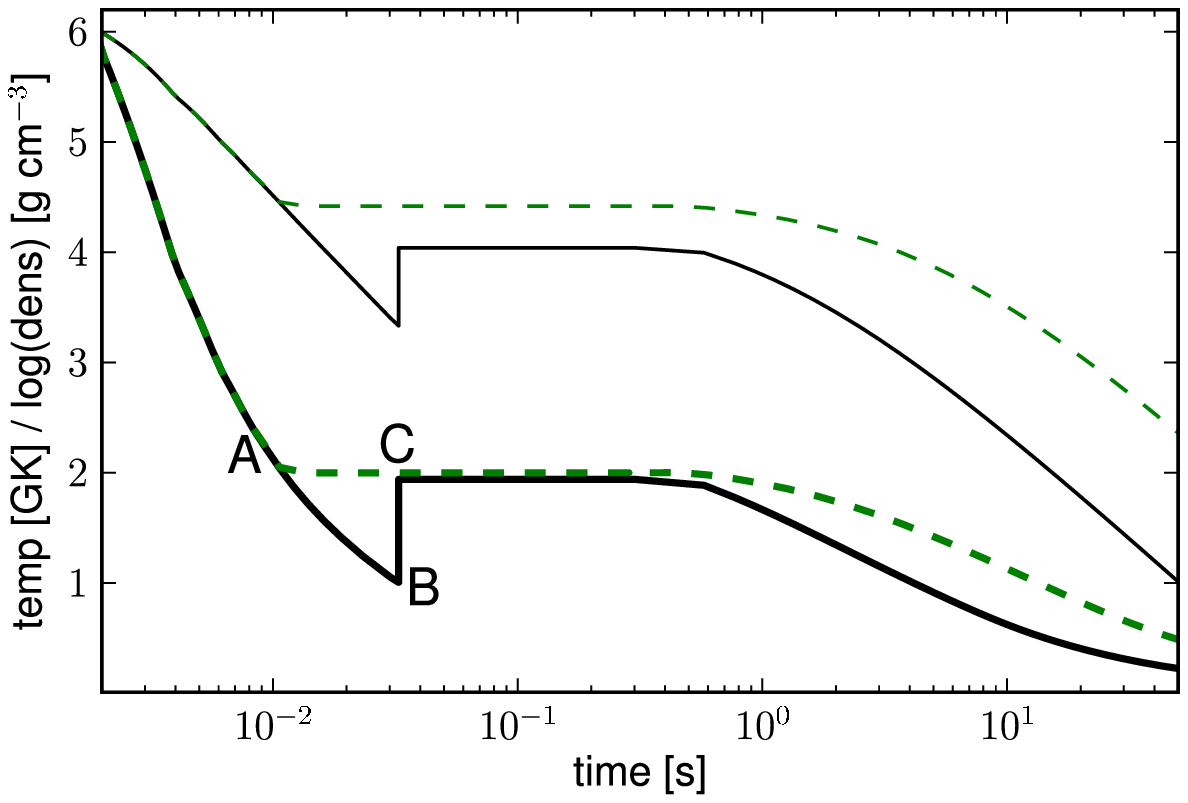}\\
 \includegraphics[width=\linewidth]{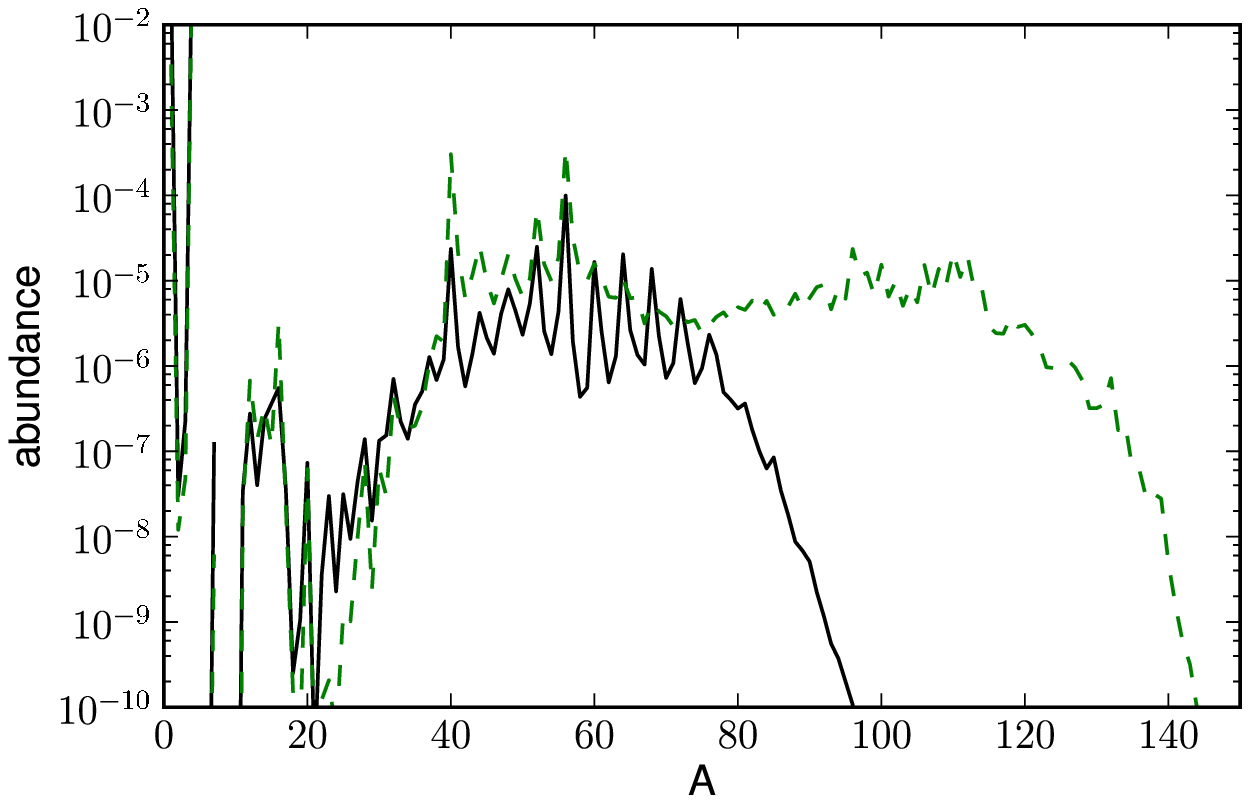}\\
 \caption{The upper panel shows the temperature (thick lines) and
   density (thin lines) evolutions with jump (solid black) and without
   jump (dashed green) at the wind termination. The corresponding
   abundances vs. mass number are presented in the lower panel.}
  \label{fig:jump}
\end{figure}

When matter moves supersonically there is a jump at the wind
termination in temperature and density, also called reverse
shock. Here, we analyze the impact of such a jump on the $\nu
p$-process nucleosynthesis using the trajectories shown in
Fig.~\ref{fig:jump}. The trajectory without jump corresponds to the
evolution with $T_{\mathrm{wt}}=2$~GK presented in the previous
section. The trajectory with jump is chosen such that the temperature
increases after the wind termination reaches $\approx 2$~GK (same
temperature as in the trajectory without jump).  The constant
temperature phase is the same for both evolutions ($\Delta t
=0.5$~s). However, the final abundances are very different as shown in
Fig.~\ref{fig:jump}.

In the evolution with jump, the expansion continues very fast between
the moment the temperature reaches 2~GK (marked with an A in the
temperature curve, Fig.~\ref{fig:jump}) and the position of the wind
termination shock (marked with B). During this phase, there is not
enough time for antineutrino absorption on protons to produce the
necessary amount of neutrons to reach heavier nuclei. Already in this
phase, between A and B, matter starts to beta decay towards stability
because temperatures become too low for proton-capture reactions. The
temperature rise at the wind termination, from point B to C, increases
the effectiveness of proton-capture reactions.  This results in the
matter flow moving again away from stability and towards heavier
nuclei.  Note that the temperature and matter distribution after the
wind termination are very similar for both trajectories (points A and
C).

\begin{figure}[!htb]
 \includegraphics[width=\linewidth]{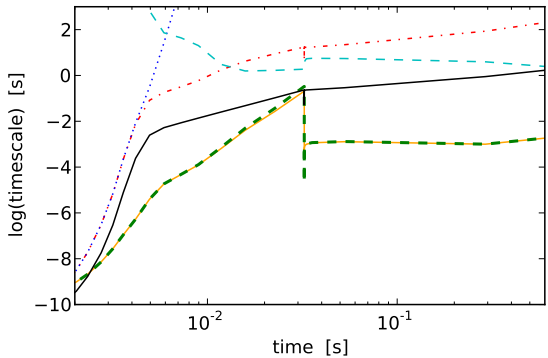}
 \caption{Same as Fig.~\ref{fig:timescale} for the evolution with a
   jump at the reverse shock.}
  \label{fig:jump_tscale}
\end{figure}

The relevant timescales for the trajectory with jump are presented in
Fig.~\ref{fig:jump_tscale}. This can be compared to the middle panel
in Fig.~\ref{fig:timescale} that corresponds to the trajectory without
jump. After the wind termination, proton-captures are
faster for the evolution without jump. More significant are the
differences in $(n,\gamma)$ and $(n,p)$ timescales: both are much
shorter in the evolution without jump (middle panel in
Fig.~\ref{fig:timescale}). These differences in the relevant
timescales have a big impact on the abundances.

\begin{figure}[!htb]
 \includegraphics[width=\linewidth]{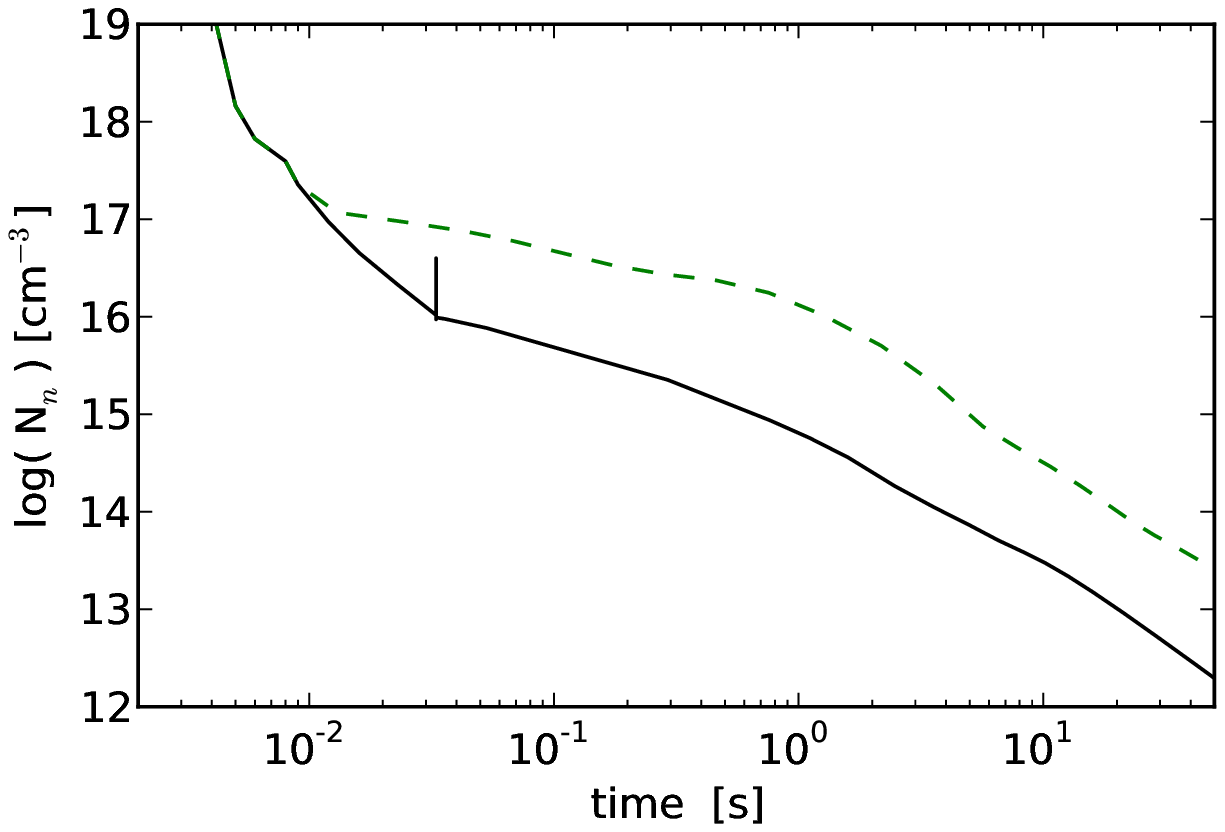}\\
 \includegraphics[width=\linewidth]{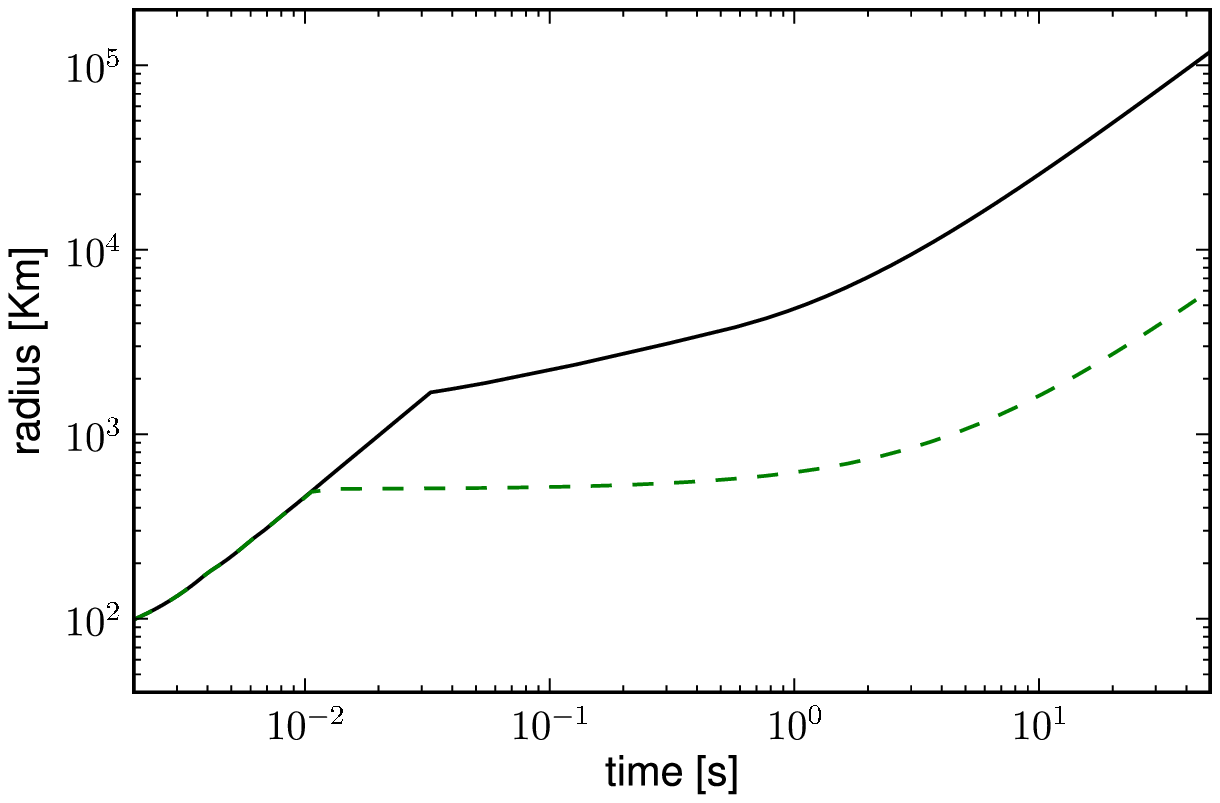}\\
 \caption{Evolution of neutron density (upper panel) and radius
   (bottom panel) for the trajectories presented in
   Fig.~\ref{fig:jump}. The solid black line corresponds to the
   evolution with jump, the dashed green line to the evolution without
   jump.}
  \label{fig:jump_Nn}
\end{figure}

The reactions involving neutrons, i.e., $(n,\gamma)$ and $(n,p)$,
depend on the neutron density which is shown in Fig.~\ref{fig:jump_Nn}
for the two evolutions.  In the trajectory with jump (solid line),
the neutron density remains lower than in the trajectory without jump
at all times, even at the wind termination (feature at $t\approx 0.02$~s).
In the $\nu p$-process there is an equilibrium between neutron capture 
and neutron production by antineutrino absorption on protons. This implies that
\begin{equation}
  \frac{\mathrm{d}Y_n}{\mathrm{d}t} = \lambda_{\bar{\nu}_e}Y_p - \sum_{Z,A} N_n Y(Z,A) \langle \sigma v \rangle_{(Z,A)} =0 .
\end{equation}
Here $\lambda_{\bar{\nu}_e}$ is the electron antineutrino absorption
rate and $\langle \sigma v \rangle_{(Z,A)}$ is the sum of reaction
rates for $(n,\gamma)$ and $(n,p)$ reactions for nucleus
(Z,~A). Therefore, the neutron density in equilibrium is given by
\begin{equation}
  N_n = \frac{\lambda_{\bar{\nu}_e}Y_p }{\sum_{Z,A}Y(Z,A) \langle \sigma v \rangle_{(Z,A)}}.
\label{eq:n-equil}
\end{equation}
The nucleosynthesis path is very similar for both trajectories
considered here (evolution with and without jump).  Therefore, only
small variations are expected in the denominator. Consequently, the
difference in the neutron densities (see Fig.~\ref{fig:jump_Nn}) is
due to the neutron production by antineutrinos.  The deceleration of
the expansion at the wind termination occurs at a smaller radii for
the trajectory without jump (see bottom panel of
Fig.~\ref{fig:jump_Nn}). The production of neutrons is hence less
efficient in the case with jump because matter reaches larger radii
where the neutrino flux is reduced due to its $r^{-2}$ dependency.
The difference in the abundance pattern for nuclei above mass number
$A \approx 80$ (see Fig.~\ref{fig:jump}) is due to this difference in
the efficiency of neutron production.

\subsection{Late-time evolution}
\label{sec:long}

We explore the impact of the dynamical evolution after the wind
termination.  In this section, we assume that the wind termination
occurs at $T_{\mathrm{wt}}=2.0$~GK and we vary the parameter $\Delta
t$.  The quantity $\Delta t$ characterizes the timescale for the
transition from an expansion with constant temperature and density
(during which $v \propto r^{-2}$) to a constant velocity expansion
with $\rho \propto r^{-2}$.  We choose the values $\Delta t
=$~0.0, 0.5, and 1.0~s, motivated by the anisotropic evolution of the
ejecta in 2D hydrodynamical simulations \cite[]{Arcones.Janka:2011}.
There is a smooth
transition between the two extreme cases of $\Delta t = 1.0$~s and
$\Delta t = 0.0$~s, which can be seen in the intermediate case of
$\Delta t = 0.5$~s. The latter ($\Delta t = 0.5$~s) was used in previous sections.

\begin{figure}[!htb]
 \includegraphics[width=0.9\linewidth]{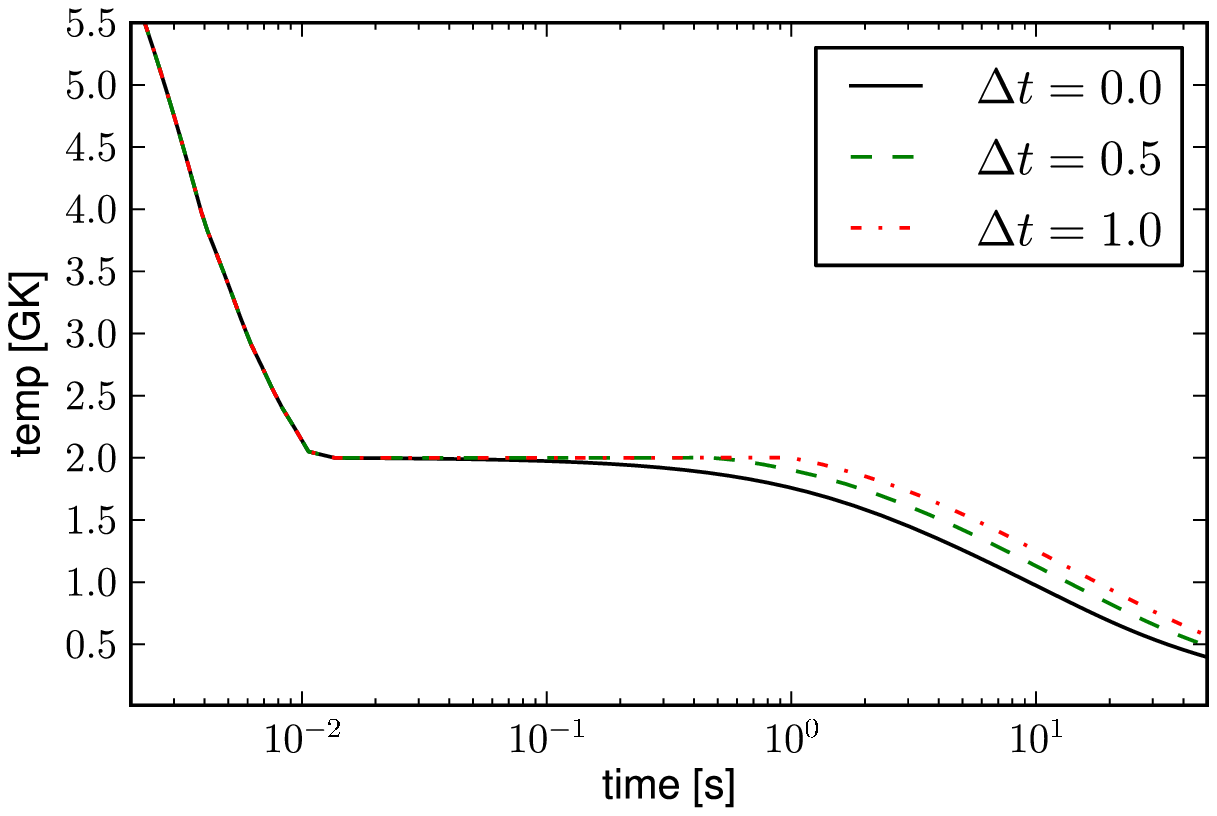}
 \includegraphics[width=0.9\linewidth]{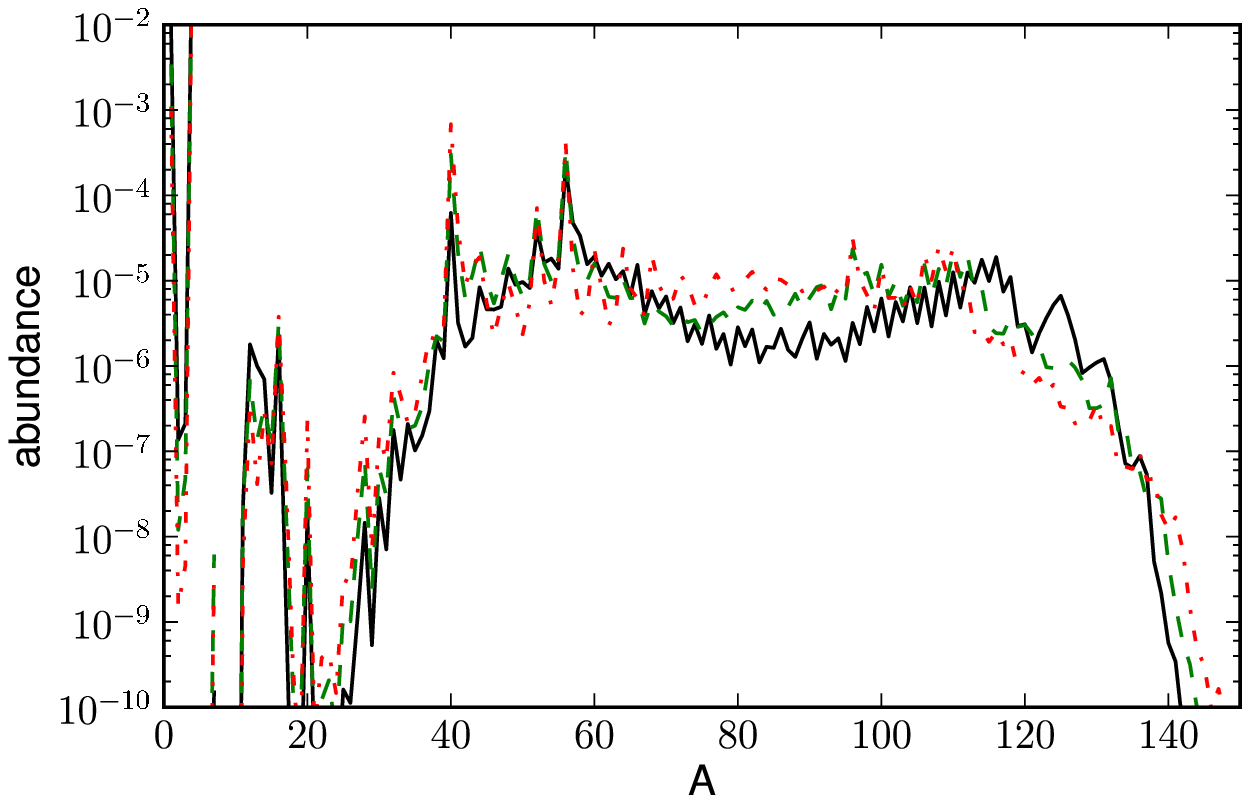}
 \includegraphics[width=0.9\linewidth]{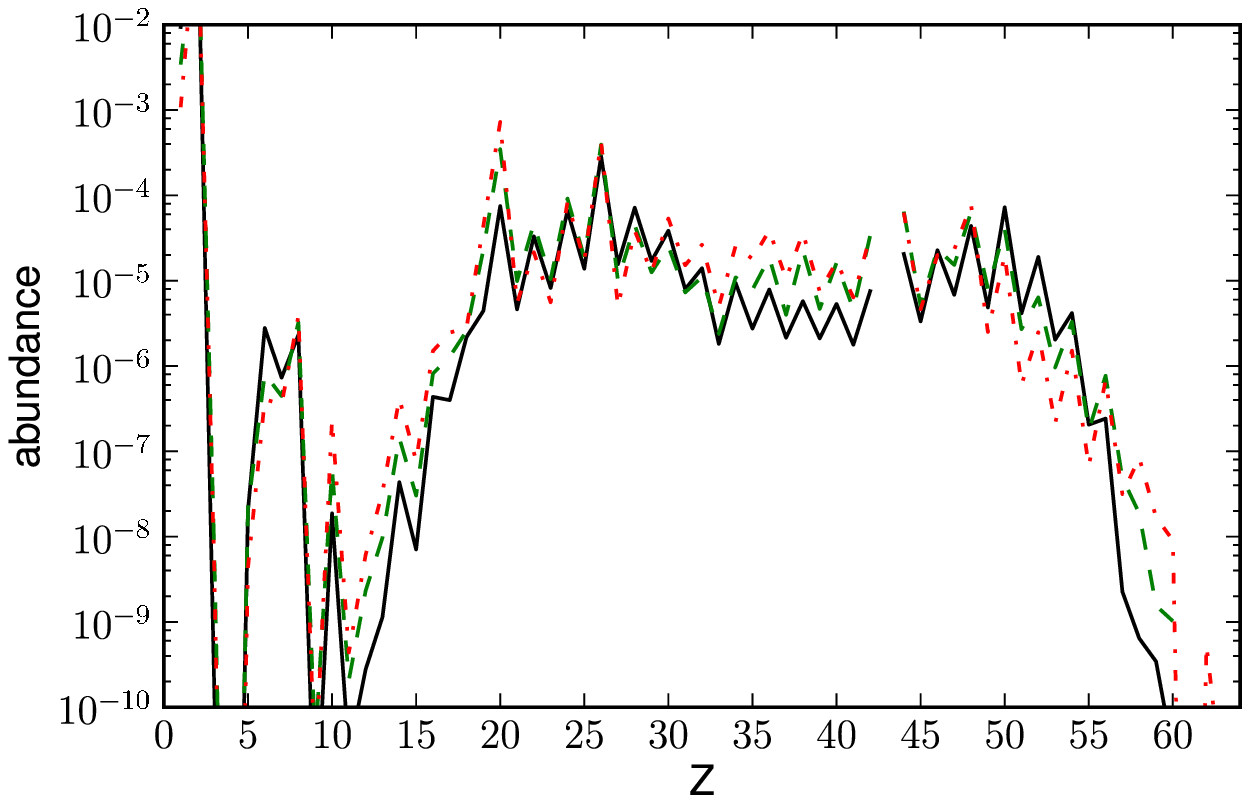}
 \caption{Different temperature evolutions after the wind termination shock
   which is at 2.0~GK are shown in the top panel.  Values for $\Delta
   t$ are 0.0~s (solid black), 0.5~s (dashed green), and 1.0~s (dotted
   red).  Abundances for these evolutions are shown versus mass number
   $A$ (middle panel) and versus atomic number $Z$ (bottom panel).}
  \label{fig:dt-Y}
\end{figure}

The different temperature evolutions (top panel) and the resulting
nucleosynthesis (middle and bottom panels) are shown in
Fig.~\ref{fig:dt-Y}.  While all three cases produce similar abundance
distributions, the details depend critically on the value of $\Delta
t$.

\begin{figure*}[!htb]
 \includegraphics[width=0.49\linewidth]{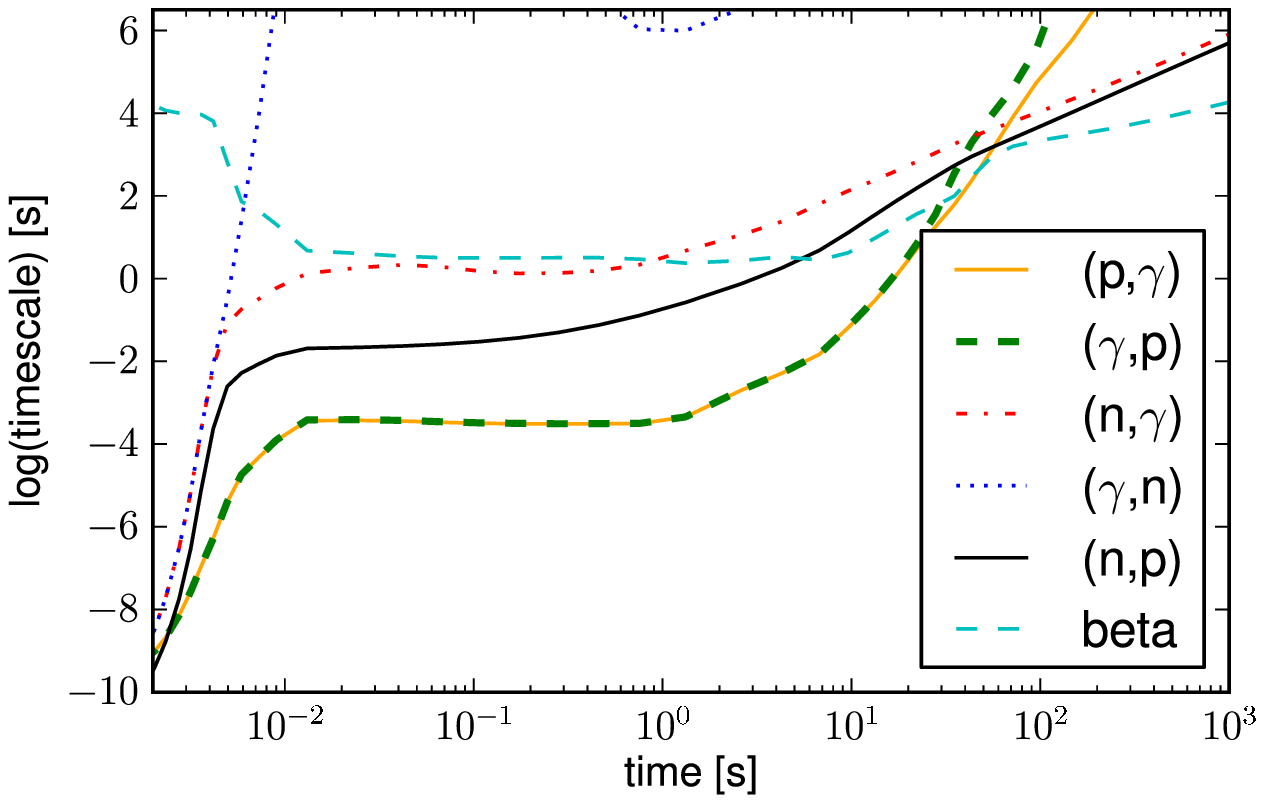} 
 \includegraphics[width=0.49\linewidth]{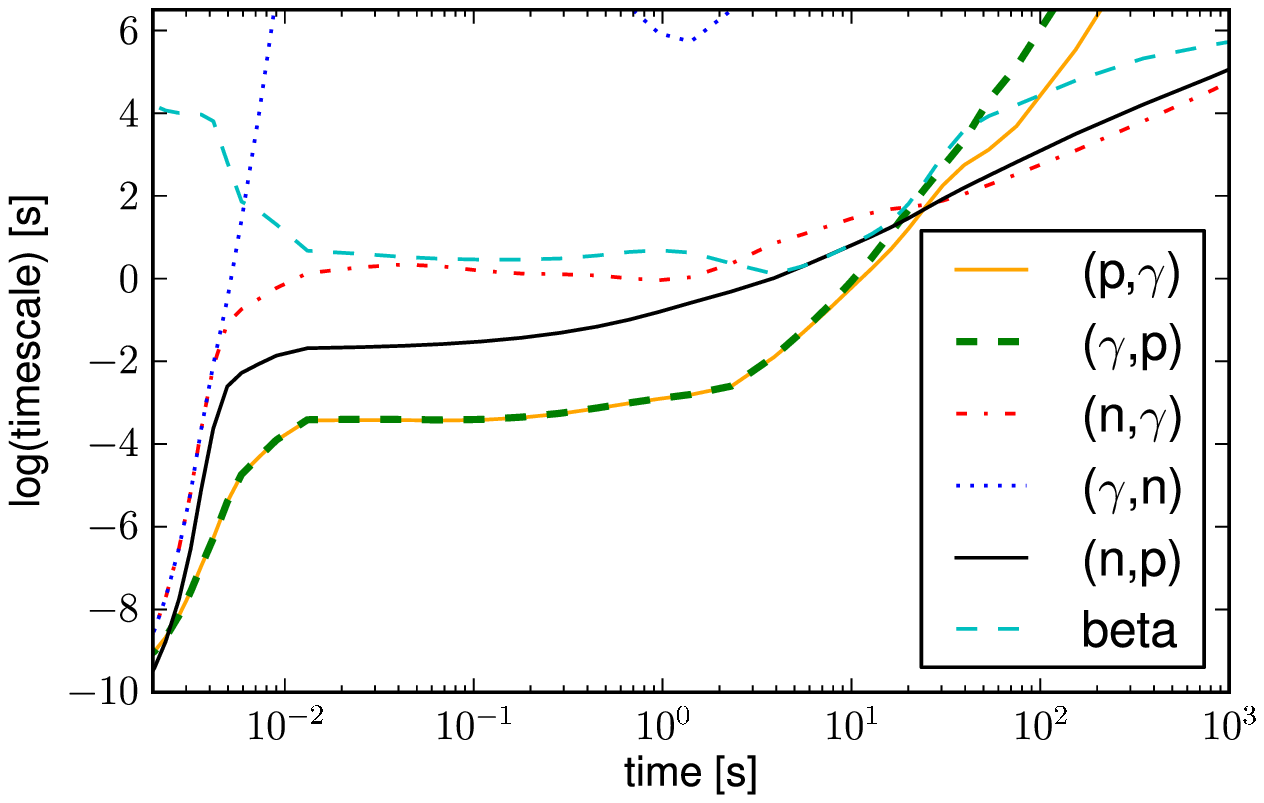} \\
 \includegraphics[width=0.49\linewidth]{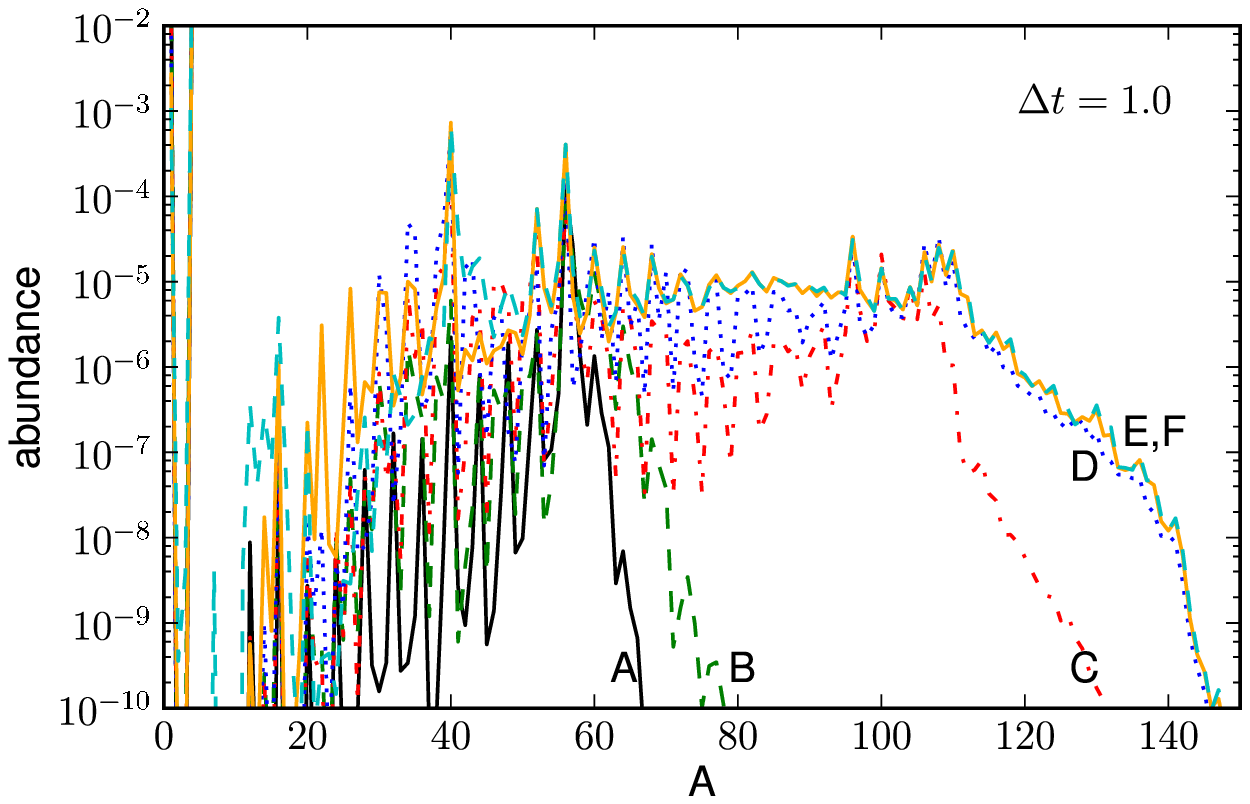}
 \includegraphics[width=0.49\linewidth]{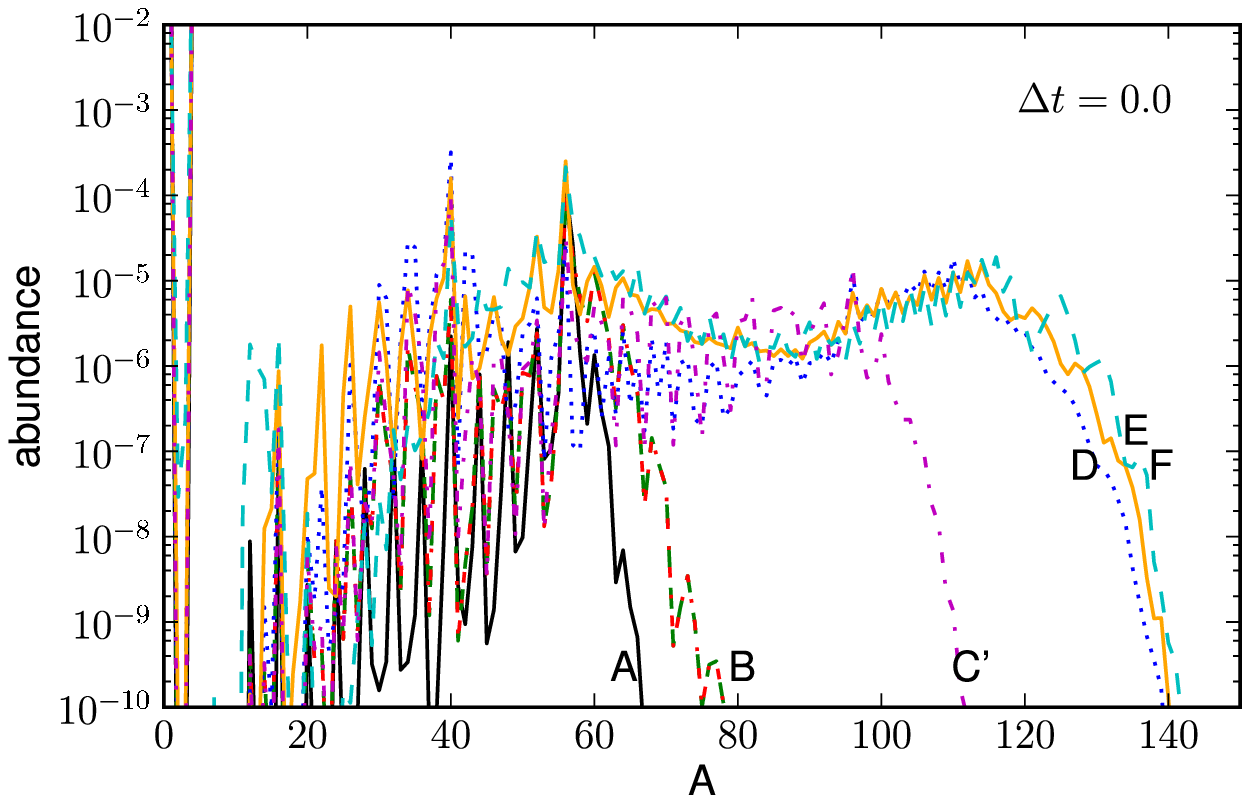} \\
 \includegraphics[width=0.49\linewidth]{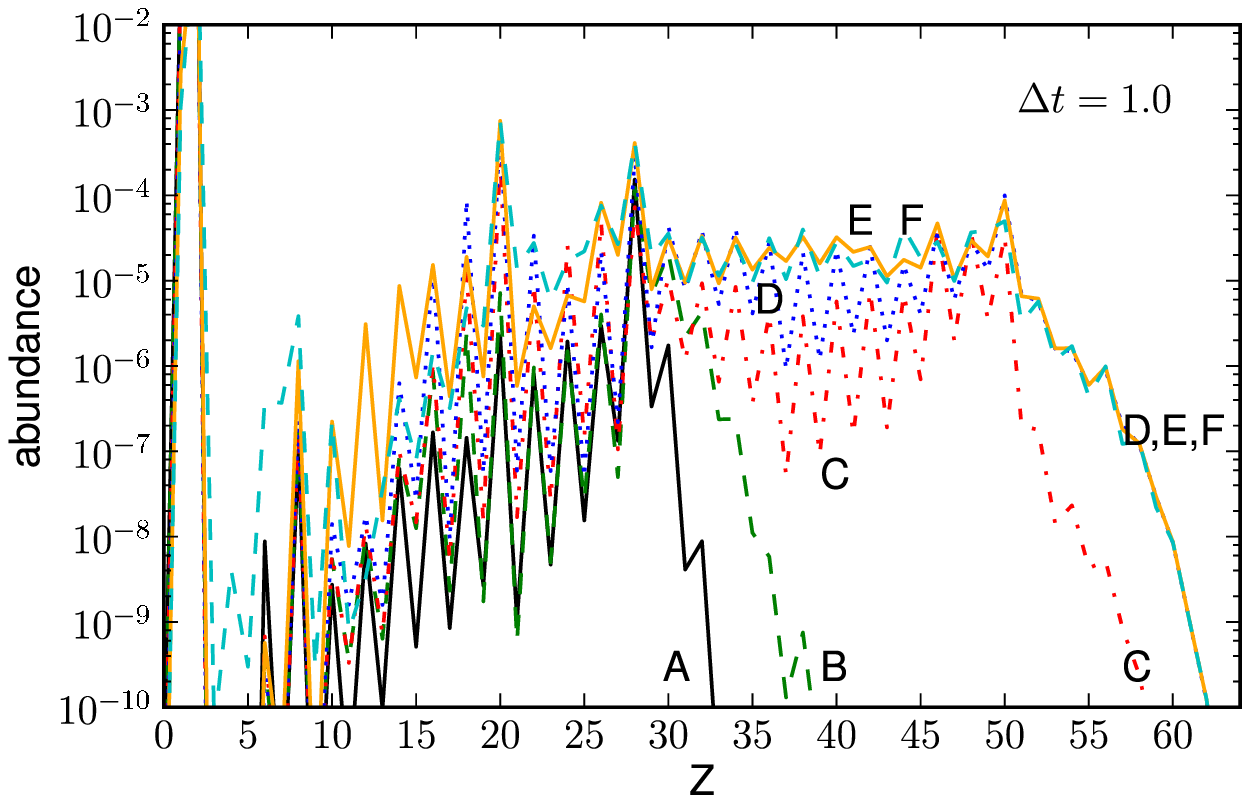}
 \includegraphics[width=0.49\linewidth]{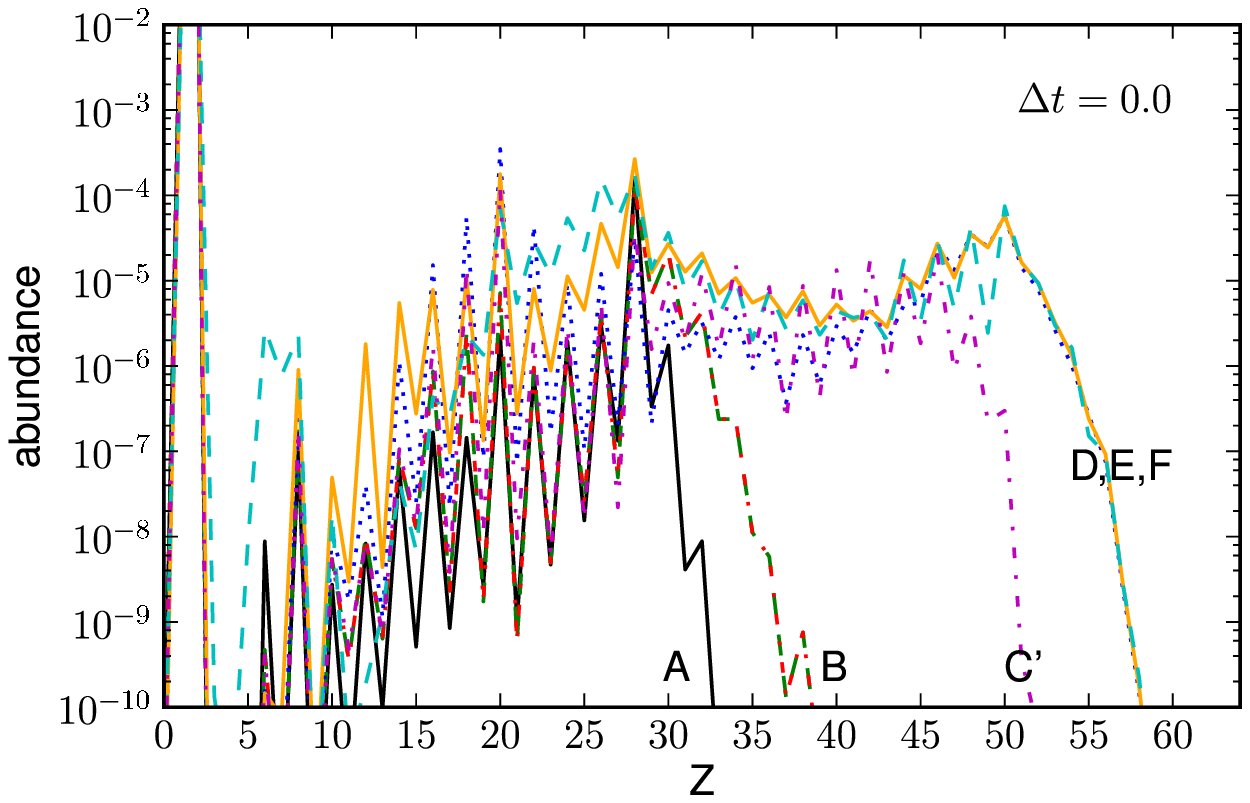} \\
 \caption{The left column is for the case of $\Delta t = 1.0$~s, the
   right column is for $\Delta t = 0.0$~s.  The top row shows average
   timescales for different reaction types as function of time.
   Abundance distributions at selected times are shown as function of
   mass number (middle row) and as function of atomic number (bottom
   row) for both cases of $\Delta t$.  For the case $\Delta t =
   1.0$~s, the abundances are shown at $T = $~3.0~(A), 2.0~(B,C),
   1.6~(D),1.0~GK~(E) and also the final abundances (F).  The
   abundances at the beginning (B) and at the end (C) of the constant
   temperature phase differ quite significantly.  For the case $\Delta
   t = 0.0$~s, the abundances are shown at $T= $~3.0~(A), 2.0~(B),
   1.9~(C'), 1.5~(D), 1.0~GK~(E) and also the final abundances (F).
   Note that while the abundances at (C) for $\Delta t = 1.0$~s and at
   (C') for $\Delta t = 0.0$~s are similar, the temperature is
   slightly lower for (C') because there is no constant temperature
   phase in this case.}
  \label{fig:dt-tau-steps}
\end{figure*}

For a relatively long phase of constant temperature ($\Delta t =
1.0$~s; left column of Fig.~\ref{fig:dt-tau-steps}), the
$(p,\gamma)-(\gamma,p)$ equilibrium lasts for quite a long time and
allows for many $(n,p)$ and $(p,\gamma)$ reactions to occur.  This
drives the matter to higher mass number (up to $A \approx 110$), where
matter accumulates in Sn ($Z=50$ closed shell).  Due to the extended
period of $(p,\gamma)-(\gamma,p)$ equilibrium, significant abundances
of nuclei with $64 \lesssim A \lesssim 110$ are synthesized (see line
C in Fig.~\ref{fig:dt-tau-steps} left column). Also during this phase
there is a continues production of seed nuclei at the expenses of
alpha particle (see Fig.11).  Once the temperature drops below
$\approx$~1.5~GK (line D), the production of seed stops. In addition,
proton-capture reactions become hindered by the Coulomb barrier and
$\beta^+$-decays become faster than $(n,p)$ reactions (see upper left
panel in Fig~\ref{fig:dt-Y}).  The nucleosynthesis ceases to
efficiently proceed to higher masses.  During this phase, matter
starts to decay towards stability.  A small number of late time
neutron capture reactions smooths the abundance distribution in the
mass range $80 \lesssim A \lesssim 110$ (line E).  Once the
temperature drops below 1.0~GK (lines E and F), the abundances as
function of mass number do not change anymore, as at this time
$\beta^+$-decays dominate.  The closed shell at $Z=50$ acts as barrier
for the nucleosynthesis.  This can be seen in the enhanced abundances
at $A=110$ (originating from $^{110,111,112}$Sn) in the final
abundance distribution.

\begin{figure}[!htb]
 \begin{center}
 \includegraphics[width=0.95\linewidth]{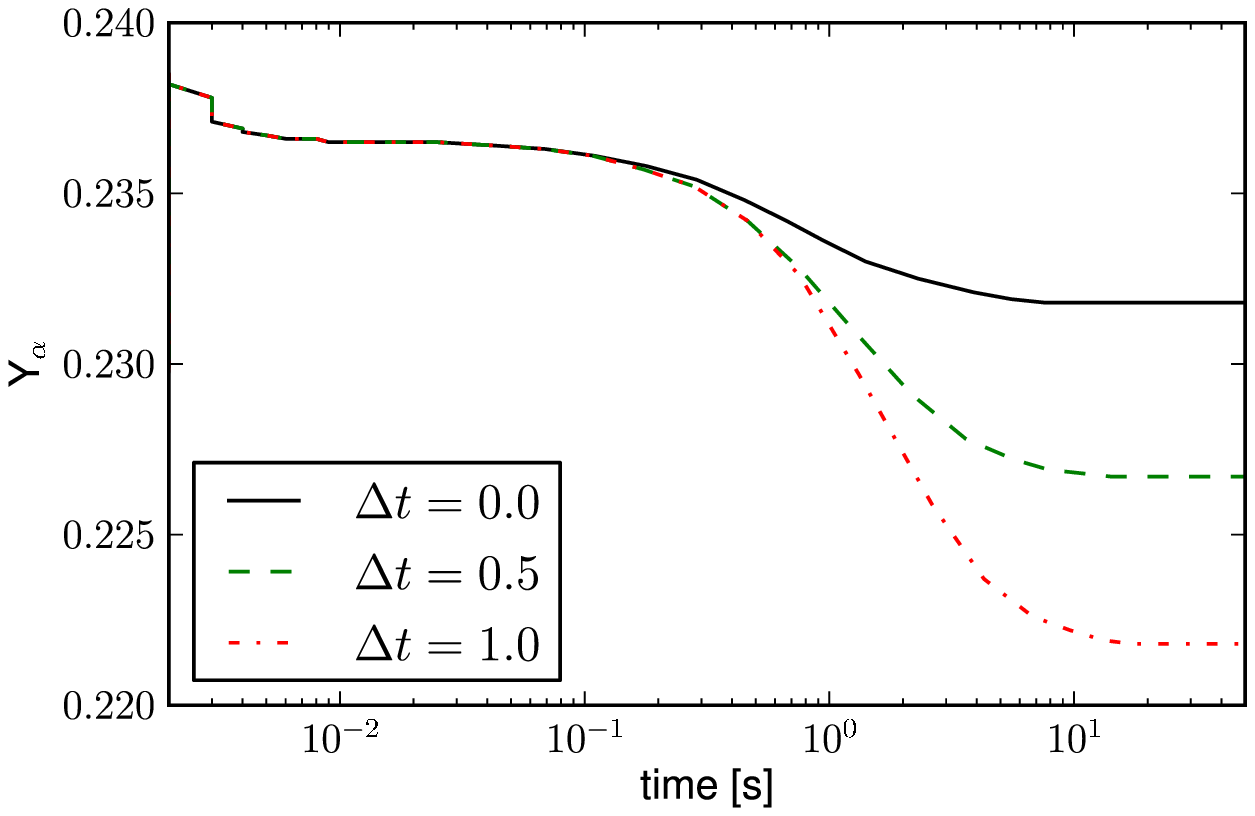}
 \includegraphics[width=0.95\linewidth]{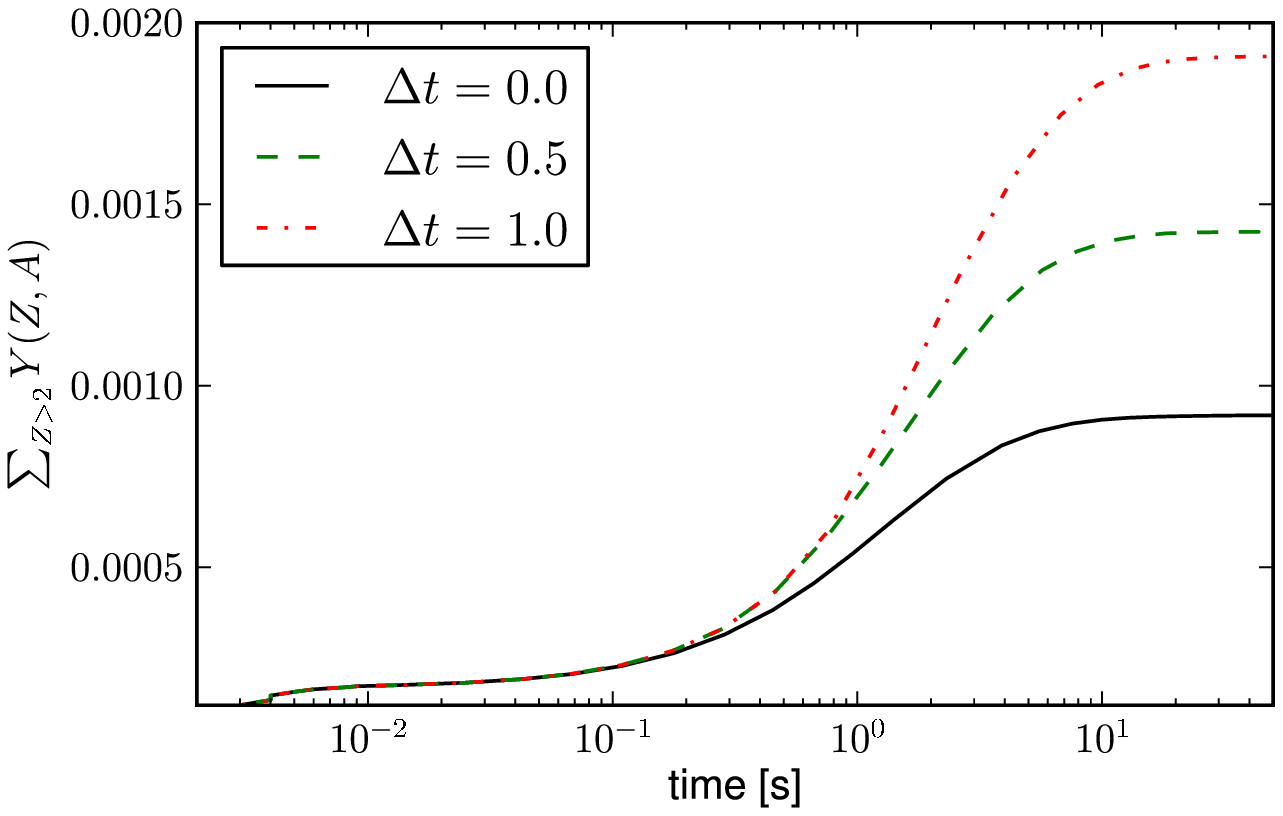}
 \includegraphics[width=0.95\linewidth]{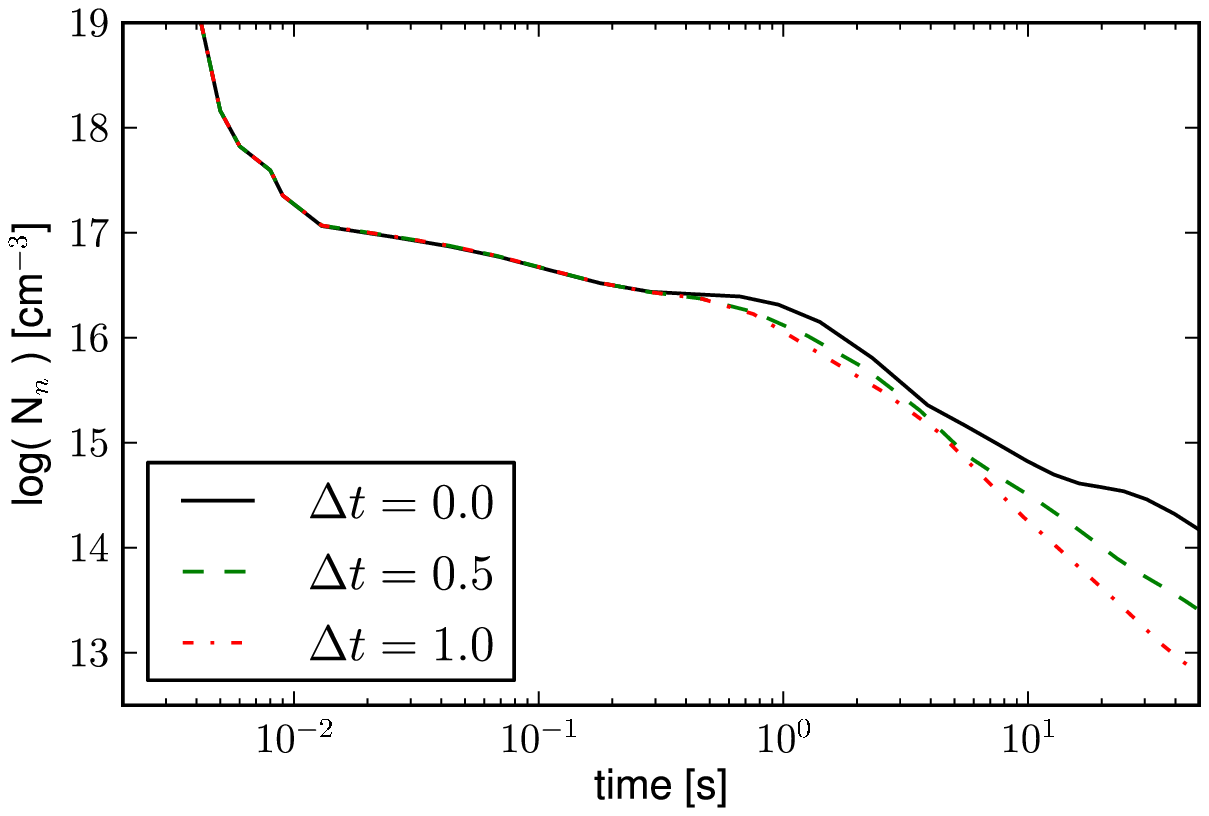}
 \end{center}
 \caption{Abundances of $\alpha$ particles (top panel), abundances of
   heavy nuclei with $Z>2$ (middle panel), and neutron densities
   (bottom panel) for $\Delta t = 0.0$~s (solid black), 0.5~s (dashed
   green), and 1.0~s (dotted red).}
 \label{fig:yn-ya-yh}
\end{figure}

For the case $\Delta t = 0.0$~s (right column in
Fig.~\ref{fig:dt-tau-steps}), the situation is very different.  The
initial phase of $(p,\gamma)-(\gamma,p)$ equilibrium is much shorter
than in the case of $\Delta t = 1.0$~s.  In this case nuclei up to
mass number $A \approx 110$ are synthesized (line C') the temperature
has already dropped below 2~GK.  Due to the faster temperature
decline, the production of seed nuclei is less efficient than in the
case of $\Delta t = 1.0$~s (see Fig.~\ref{fig:yn-ya-yh}).  This leads
to lower abundances of nuclei such as $^{12}$C, $^{16}$O, $^{20}$Ne,
$^{28}$Si, and $^{40}$Ca for $\Delta t = 0$~s (see
Fig.~\ref{fig:dt-Y}). The lower abundance of seed nuclei leads to a
higher neutron density (see Eq.~\ref{eq:n-equil}) as in both cases the
neutron production rate ( $\lambda_{\bar{\nu}_e}Y_p$) is
similar. Therefore, in the evolution with $\Delta t = 0$~s, there are
more neutrons available per seed nuclei (Fig.~\ref{fig:yn-ya-yh}).
The neutron capture reactions, i.e.  $(n,\gamma)$ and $(n,p)$, become
faster than $(p,\gamma)$ reactions at temperatures of $T < 1.5$~GK.
These neutron captures move matter to the neutron-rich side of
stability.  This, together with the earlier less efficient production
of seeds, results in a depletion of the abundances in the region of
$64 \lesssim A \lesssim 110$ (compare black solid and red dashed line
in Fig.~\ref{fig:dt-Y}).  Towards the end of the evolution,
$(n,\gamma)$ reactions are the fastest ones, even faster than
$\beta^+$-decays.  During this phase, the atomic number $Z$ remains
unchanged while the mass number increases (lines D, E, and F in
Fig.~\ref{fig:dt-tau-steps}).  The signature of matter moving to
stability via $(n,\gamma)$ reactions can be seen in the increased
abundances at $A = 124$, corresponding to $^{124}$Sn, the most
neutron-rich stable Sn isotope.

\section{Conclusions}
\label{sec:conclusions}

We have studied the impact of the supernova dynamical evolution on the
$\nu p$-process and how the wind termination affects the synthesis of
elements beyond iron. For a robust and strong production of nuclei
with $64 \lesssim A \lesssim 110$ there is an optimal wind termination
temperature at $\approx$2~GK as it was found by
\cite{Wanajo.etal:2011}. Also in agreement with their work, we have
shown that if the wind termination occurs at low temperatures
($T_{\mathrm{wt}}\approx 1$~GK), matter stays too short time in the
optimal temperature range for $\nu p$-process nucleosynthesis, $\sim 1
-2 $~GK. Moreover, the electron antineutrino flux quickly becomes too
small to produce the necessary neutrons to overcome the
$\beta^{+}$-decay waiting points. This hinders the $\nu p$-process and
consequently the efficient synthesis of heavy nuclei.  Therefore, the
wind termination temperature determines the heaviest elements
produced.

We have identified an end point cycle that is key at high temperatures
(around 3~GK). The close NiCu cycle inhibits the production of elements 
heavier than 56Ni. The reactions in this closed NiCu cycle are the
following:
\begin{eqnarray}
&& ^{56}Ni (n,p) ^{56}Co (p, \gamma ) ^{57}Ni (n,p) ^{57}Co (p, \gamma ) ^{58}Ni \nonumber \\
& &\longrightarrow \, ^{58}Ni (p, \gamma ) ^{59} Cu(p, \alpha ) ^{56} Ni \, . \nonumber
\end{eqnarray}
At high temperatures the reaction $^{59}$Cu$(p,\alpha)$ $^{56}$Ni
prevents the synthesis of elements beyond the iron group. When the
temperature drops below $\approx 3$~GK, $^{59}$Cu$(p,\gamma)$
$^{60}$Zn becomes more effective than $^{59}$Cu$(p,\alpha)$
$^{56}$Ni.  This leads to breakout from the NiCu cycle and the flow of
matter can continue towards heavier nuclei.  The temperature
dependence of these two reactions is critical because it sets the
temperature at which the $\nu p$-process can start to synthesize
elements beyond Nickel. If this cross-over temperature were slightly
higher, matter would be closer to the proto-neutron star and thus
under higher neutrino flux when the path starts to move towards
heavier nuclei. This will significantly increase the efficiency of the
$\nu p$-process. Therefore, our results clearly motivate further
investigation of these two key reactions: $^{59}$Cu$(p,\alpha)$
$^{56}$Ni and $^{59}$Cu$(p,\gamma)$$^{60}$Zn. Particularly relevant
are the branching ratios for the decay by alpha and gamma emission of
compound states in $^{60}$Zn above the proton separation energy.

We have also explored for the first time the impact of the dynamical
evolution after the wind termination on the synthesis of heavy
elements by the $\nu p$-process. Hydrodynamical simulations show that
after the wind termination there is a transition from an expansion with
almost constant temperature and density to a phase with almost
constant velocity. The duration of the constant temperature phase,
$\Delta t$, strongly affects the final abundances of heavy nuclei. We
have found that for $\Delta t=0.0$~s the final flow of matter to
stability occurs by neutron capture reactions, i.e\ $(n,\gamma)$ and
$(n,p)$, and not by $\beta^+$-decays.  When the phase of constant
temperature is very short ($\Delta t < 0.5$~s), the abundances of
nuclei with $64<A<110$ are lower.  This leads to significantly higher
neutron densities at later times.  Depending on the neutron density
the matter moves to stability either by $\beta^+$-decays ($\Delta t =
1.0$~s) or by neutron captures.  These two different ways of reaching
stability leave a distinct fingerprint in the final abundances.

We have investigated in detail the impact of dynamical evolution on 
the $\nu p$-process nucleosynthesis using individual characteristic
trajectories. In order to predict the complete $\nu p$-process
yields from a supernova simulation, one will need to integrate 
over all proton-rich ejecta. 

In summary, the supernova dynamics as well as individual reactions 
such as the proton capture reactions on $^{59}$Cu determine how
high in proton number the $\nu p$-process can proceed. The dynamical 
evolution just after the wind termination can
greatly affect the nucleosynthesis evolution towards stability as it
determines the late neutron density.  Our results provide a link
between nucleosynthesis in proton-rich winds and the dynamical
evolution of the ejected matter and motivate further theoretical and
experimental effort on understanding key reactions.

\acknowledgments

We thank F.~Montes, T.~Rauscher, and F.-K.~Thielemann for valuable
discussions and U.~Frischknecht and C.~Winteler for their support
preparing the flux figures. A.A.~acknowledge support from the
Alexander von Humboldt Foundation and the Swiss National Science
Foundation. C.F.~acknowledges support from the DOE Topical
Collaboration "Neutrinos and Nucleosynthesis in Hot and Dense Matter"
under contract DE-FG02-10ER41677. G.M.P. is partly supported by the
Deutsche Forschungsgemeinschaft through contract SFB 634, the
Helmholtz International Center for FAIR within the framework of the
LOEWE program launched by the state of Hessen and the Helmholtz
Association through the Nuclear Astrophysics Virtual Institute
(VH-VI-417).


\end{document}